\documentclass[final,3p,times]{elsarticle}
\usepackage{amssymb}
\usepackage{amsthm}

\usepackage{lineno}
\usepackage{natbib} 

\usepackage{hyperref}
\usepackage{url}

\usepackage{amsmath,amsfonts}
\usepackage{algorithmic}
\usepackage{algorithm}
\usepackage{array}
\usepackage[caption=false,font=normalsize,labelfont=sf,textfont=sf]{subfig}
\usepackage{textcomp}
\usepackage{stfloats}
\usepackage{verbatim}
\usepackage{graphicx}

\usepackage{multirow}
\usepackage{soul}
\usepackage[table]{xcolor}
\usepackage{times}
\usepackage{epsfig}
\usepackage{graphicx}
\usepackage{amssymb}
\usepackage{soul}
\usepackage{enumitem}
\usepackage{verbatim}
\usepackage{multirow}
\usepackage{float}
\usepackage[export]{adjustbox}
\usepackage{textcomp,array,booktabs}

\usepackage{wrapfig,lipsum,booktabs}

\usepackage{mathtools}
\usepackage{booktabs}% http://ctan.org/pkg/booktabs

\DeclareUnicodeCharacter{2212}{-}

\usepackage{multirow}
\usepackage{makecell}
\usepackage{multicol,caption}

\journal{Engineering Applications of Artificial Intelligence}

\newcommand{\norm}[1]{\| #1 \|}
\newcommand{\bignorm}[1]{\Bigl \| #1 \Bigr \| }

\begin{document}

\begin{frontmatter}

\title{Topological Persistence Guided Knowledge Distillation \\ for Wearable Sensor Data}

\author[label1]{Eun~Som~Jeon}
\author[label1]{Hongjun~Choi}
\author[label1]{Ankita~Shukla}

\affiliation[label1]{organization={Geometric Media Lab, School of Arts, Media and Engineering and School of Electrical, Computer and Energy Engineering, Arizona State University},
          %addressline={},
             city={Tempe},
             postcode={85281},
             state={AZ},
             country={USA}}

\author[label2]{Yuan Wang}
\affiliation[label2]{organization={Department of Epidemiology and Biostatistics, University of South Carolina},
          %addressline={},
             city={Columbia},
             postcode={29208},
             state={SC},
             country={USA}}
\author[label3]{Hyunglae Lee}
\affiliation[label3]{organization={School for Engineering of Matter, Transport and Energy},
          %addressline={},
             city={Tempe},
             postcode={85281},
             state={AZ},
             country={USA}}
\author[label4]{Matthew~P.~Buman}
\affiliation[label4]{organization={College of Health Solutions, Arizona State University},
          %addressline={},
             city={Phoenix},
             postcode={85004},
             state={AZ},
             country={USA}}

\author[label1]{and~Pavan~Turaga}

\begin{abstract}
Deep learning methods have achieved a lot of success in various applications involving converting wearable sensor data to actionable health insights. A common application areas is activity recognition, where deep-learning methods still suffer from limitations such as sensitivity to signal quality, sensor characteristic variations, and variability between subjects. To mitigate these issues, robust features obtained by topological data analysis (TDA) have been suggested as a potential solution. 
However, there are two significant obstacles to using topological features in deep learning: (1) large computational load to extract topological features using TDA, and (2) different signal representations obtained from deep learning and TDA which makes fusion difficult.
In this paper, to enable integration of the strengths of topological methods in deep-learning for time-series data, we propose to use two teacher networks -- one trained on the raw time-series data, and another trained on persistence images generated by TDA methods. These two teachers are jointly used to distill a single student model, which utilizes only the raw time-series data at test-time.
This approach addresses both issues. The use of KD with multiple teachers utilizes complementary information, and results in a compact model with strong supervisory features and an integrated richer representation.
To assimilate desirable information from different modalities, we design new constraints, including orthogonality imposed on feature correlation maps for improving feature expressiveness and allowing the student to easily learn from the teacher.
Also, we apply an annealing strategy in KD for fast saturation and better accommodation from different features, while the knowledge gap between the teachers and student is reduced.
Finally, a robust student model is distilled, which can at test-time uses only the time-series data as an input, while implicitly preserving topological features. The experimental results demonstrate the effectiveness of the proposed method on wearable sensor data. The proposed method shows 71.74\% in classification accuracy on GENEActiv with WRN16-1 (1D CNNs) student, which outperforms baselines and takes much less processing time (less than 17 sec) than teachers on 6k testing samples.

\end{abstract}


\end{frontmatter}



\section{Introduction}
Wearable sensor data, used with deep learning methods, has achieved great success in various fields such as smart homes, health-care services, and intelligent surveillance \cite{nweke2018deep}. %However, analysis of wearable sensor data suffers from particular challenges because of inter- and intra-person variability, and dependency on segmentation method and sampling rate for analysis \cite{seversky2016time}.
However, analysis of wearable sensor data suffers from particular challenges because of inter- and intra-person variability and noisy signal problems \cite{seversky2016time, edelsbrunner2022computational}. To mitigate these problems, utilizing robust features obtained by methods such as topological data analysis (TDA) have been proposed as a solution, and has proven beneficial \cite{seversky2016time}. TDA in fusion with machine learning methods has achieved significant results in stock market analysis \cite{gholizadeh2018short, yen2021using}, time-series forecasting \cite{zeng2021topological}, disease classification \cite{pachauri2011topology, nawar2020topological}, and texture classification \cite{edelsbrunner2022computational}.

TDA has been used to characterize the shape of complex data, using the persistence of connected components and high-dimensional holes which are computed by the persistent homology (PH) algorithm  \cite{edelsbrunner2022computational}. The persistence information can be represented by features such as persistence image (PI) \cite{adams2017persistence}. However, two key challenges are commonly reported in utilizing topological features: (1) the large computational, memory, and runtime requirements to extract persistence features from large-scale data \cite{som2020pi} make it challenging to implement on small devices with limited computational power. Utilizing two modalities also requires an increase in the computational power to store and interpret the data. (2) Further, a significant compatibility gap between the univariate signal data and datastructures from TDA feature representations make it difficult to integrate them in a unified framework. These differences in feature representations make conventional models difficult to use for fusing these different representations.

%Based on these observations, in this paper, we propose a new framework using two teachers in knowledge distillation with the time-series and persistence images \es{to generate a better student utilizing the original time-series data only as an input.}

Knowledge distillation (KD) has been utilized to generate a smaller model (student) from the learned knowledge of a larger model (teacher) \cite{hinton2015distilling}. It has been demonstrated to have outstanding performance in the analysis of wearable sensor data \cite{jeon2022kd, som2020pi, gou2021knowledge}.
Also, using multiple teachers in KD has been studied to provide richer information, which is generally implemented with uni-modal data \cite{reich2020ensemble, liu2020adaptive, gou2021knowledge}. %

\begin{figure*}[!t]
\centering
\includegraphics[width=6.4in]{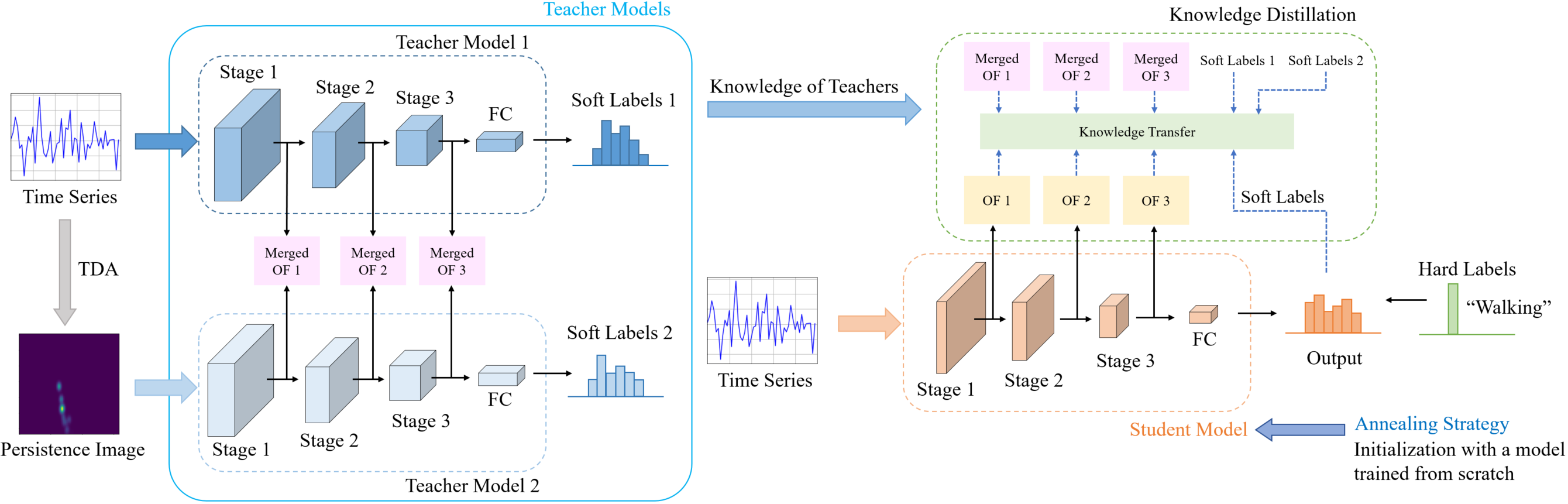}
\caption{An overview of Topological Persistence Guided Knowledge Distillation (TPKD). Two teachers, learned with different representations of the same raw time-series data, are utilized to train a compact student model.}
\label{figure:framework}
\end{figure*}

In this paper, we propose a new framework based in knowledge distillation, which enables a single compact model to be developed, that implicitly fuses the strengths of the different representations as represented by two different teachers. The distilled student is shown to acquire benefits from both teachers trained with different modalities -- raw time-series and persistence image from TDA. 
We term our approach Topological Persistence Guided Knowledge Distillation (TPKD). An overview of the TPKD is presented in Figure \ref{figure:framework}. %As shown in the figure, we use two teacher models to provide richer information, concurrently utilizing orthogonal features (OF) from correlation maps of intermediate layers, and an annealing strategy is applied to reduce the knowledge gap and to consider inherent features of the student model. More specifically, 
As seen in the figure, firstly, we extract PIs from persistence diagrams with TDA. We then train two models with time-series data and PIs, respectively. Secondly, we use the two pre-trained models as teachers separately in KD. The features from intermediate layers are transformed to a correlation map, reflecting the similarities of samples for a mini-batch in the activations of the network, and the maps from teachers are merged for integrating the features for distillation. %To provide a better quality of knowledge to transfer, orthogonal features (OF), containing more attentive relationship between features, are generated from a fused map by the activation maps of teachers.
However, since features are from different modalities, it is not clear if fusion can be done simply at the activation map level \cite{gou2021knowledge}. To better accommodate information from different modalities, we construct a new form of knowledge utilizing orthogonal features (OF), representing prominent relationships between features in multimodal streams \cite{wang2020orthogonal, choi2020role}. Orthogonality between feature-maps acts a proxy to create disentangled representations \cite{ShuklaBMVC2019}. Based on these orthogonal features, we hypothesize that a student can more easily learn from teachers of different strengths.
In the third step, to reduce the knowledge gap and consider the properties inherent in the model using time-series as an input, we apply an annealing strategy in KD. The annealing strategy guides the student model to initialize its weights from a model learned from scratch, instead of random initialization. Finally, a robust and small model is distilled by the proposed method, which uses the raw time-series data only as its input.

The contributions of this paper are as follows:
\begin{itemize}
\item We propose a new framework based on knowledge distillation that transfers topological features to the student using time-series data only as an input.
\item We develop a technique for leveraging orthogonal features from intermediate layers and an annealing strategy in KD with multiple teachers, which reduces the statistical gap in features between teachers and student for better knowledge transfer. 
\item We show strong empirical results demonstrating the strength of our approach with various teacher-student combinations on wearable sensor data for human activity recognition.
\end{itemize}

The rest of the paper is organized as follows. In section \ref{sec:background}, we provide a brief overview of generating PIs, KD techniques, and an annealing strategy. In section \ref{sec:proposed_method}, we introduce the proposed method, a new framework in KD. In section \ref{sec:experiments}, we describe our experimental results and analysis. In section \ref{sec:conclusion}, we discuss our findings and conclusions.

% \es{introduce wearable sensor data and challenge}

% \es{Topological features}

% %\es{reason why use KD and difficulties in using two different modality}

% \es{we observed two key challenges -- Extracting PI takes large time / multiple teachers with different modality generate knowledge gap}

% \es{so we propose a new framework in Topological Feature Preserving KD}

\section{Background} \label{sec:background}

%\subsection{Persistence Images}
\subsection{Topological Feature Extraction}

TDA has been utilized in various fields \cite{adams2017persistence, WANG2021109324, gholizadeh2018short,Munch2017}, achieving many successes in providing novel insights about the `shape' of data. These features have been found useful in machine learning pipelines for different applications \cite{gholizadeh2018short, zeng2021topological, som2020pi, Krim2016}.
As a key algorithm of TDA, persistent homology tracks the variations in $n$-dimensional holes present in data, characterized by points, edges, and triangles by a dynamic thresholding process, which is called a filtration \cite{edelsbrunner2002}. The persistence of these topological cavities during a filtration is described in a data-structure, such as a persistence diagram (PD) which encodes the birth and death times as $x$ and $y$ coordinates of planar scatter points \cite{adams2017persistence, edelsbrunner2022computational}.
Utilizing PDs directly in machine learning is challenging because of their heterogeneous nature, implying that the number and locations of the scatter points are not fixed and can be different due to slight perturbations of the underlying data. Organizing the scatter points based on their persistence (life time) provides a way to vectorize the PDs.

\begin{figure}[htb!] %t!
%\begin{tabular}
\includegraphics[width = 0.43\textwidth]
{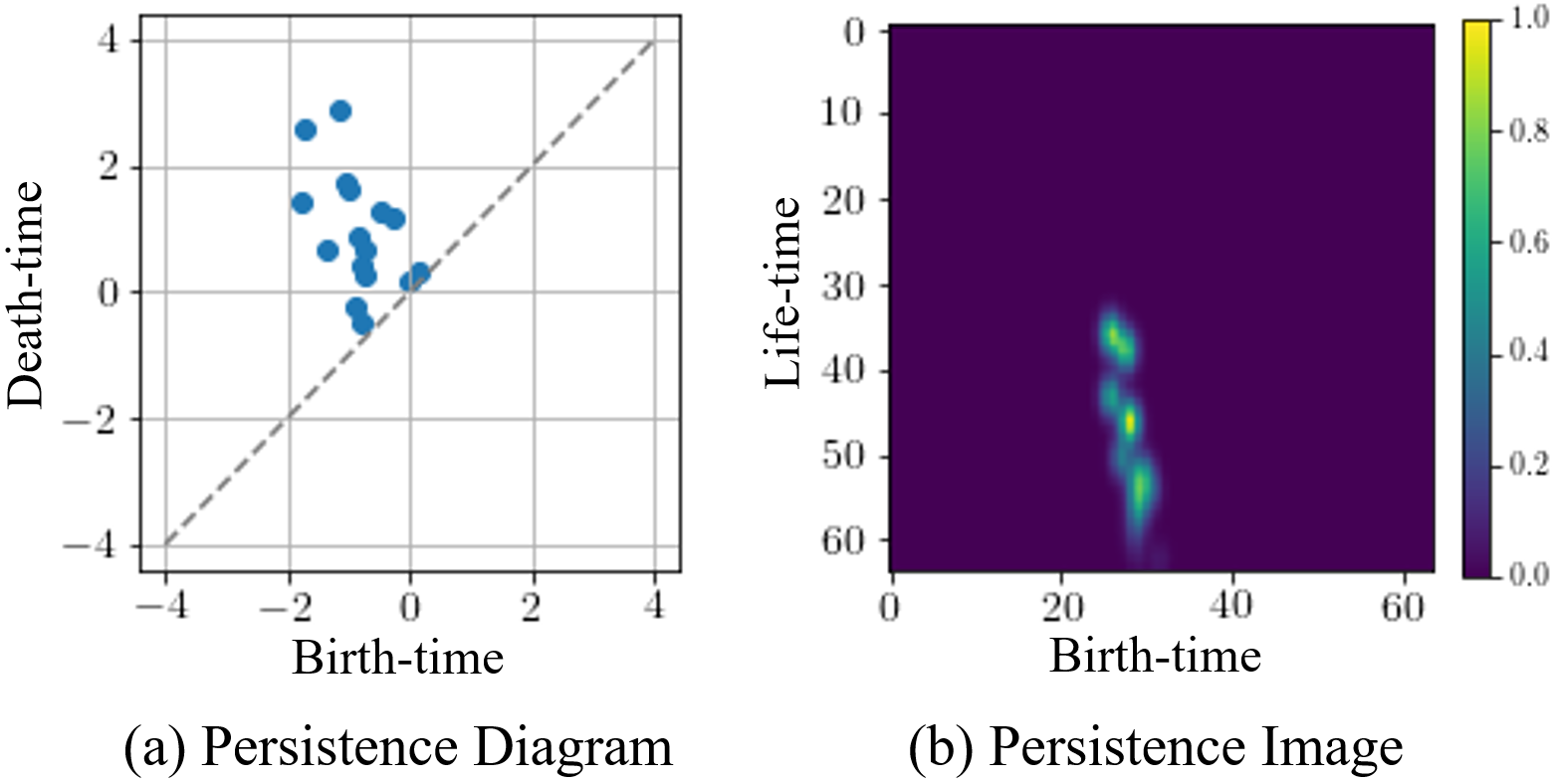} %0.27
\centering
\caption{PD and its corresponding PI. In PD, higher life-time appears brighter.}
%\end{tabular}
\label{figure:PD_PI}
\end{figure}

Persistence image (PI) is a vector representation of the PD, which represents the lifetime of homological structures in data. Firstly, to construct the PI, PD is projected into a persistence surface (PS) $\rho: \mathbb{R} \rightarrow \mathbb{R}^2$, defined by a weighted sum of Gaussian functions centered at the scatter points in the PD. The PS is discretized and results in a grid. By integrating the PS over the grid, PI is obtained and represented as a matrix of pixel values. The higher values of a PI imply high-persistence points of the corresponding PD. An example of a PD and its corresponding PI are depicted in Figure \ref{figure:PD_PI}.
Even if TDA can provide complementary information to improve the performance, since extracting PIs by TDA requires large memory and time consumption \cite{som2020pi}, it is difficult to implement the method on small devices with limited computational resources. %Also, leveraging two different modalities requires to increase computational power and memory to generate and store the data in test-time.
To solve this issue, we propose a method in knowledge distillation. We utilize features from topological time-series analysis to distill a smaller model that generates good performance as a larger model.

%%%%%%%%%%%%%%%%%%%%%%%%%%%%

%%%%%%%%%%%%%%%%%%%%%%%%%%%%%%%%%%%%%
\subsection{Deep Learning for Activity Recognition}
Deep learning methods have been broadly adopted to overcome limitations for human activity recognition (HAR) tasks, where the challenges are heuristic feature extraction relying on human experience and analyzing high-level activities \cite{wang2019deep}. In deep learning models, feature extraction and model building procedures are conducted simultaneously. In general, convolutional neural network (CNN), autoencoder, and recurrent neural network (RNN) with long-short term memory (LSTM) are utilized to build deep learning models. CNN has the advantages of analyzing relationships or correlations for nearby signals and invariance.
Recently, stacked autoencoder (SAE) \cite{zheng2016exploiting, wang2016human} was introduced to make a deeper model and to learn a better latent representation for activity classification, which is the stack of some autoencoders.
Recurrent neural network (RNN) with long-short term memory (LSTM) are also utilized popularly for HAR tasks \cite{edel2016binarized, hammerla2016deep}, which uses the temporal correlations between neurons. Furthermore, combining models, such as CNN with RNN, can enhance the ability to learn more knowledge and to recognize different activities \cite{singh2017transforming}.
Even though several different deep learning models can extract features better than heuristic methods for HAR tasks, most strategies focus on staking more layers to improve classification ability, which increases computational time and resources. The requirements are major limitations for use on small devices or real-time systems.
To address these issues, many techniques have been explored, such as network pruning \cite{molchanov2016pruning, han2015deep}, quantization \cite{wu2016quantized}, low-rank factorization \cite{tai2015convolutional}, and knowledge distillation (KD) \cite{hinton2015distilling}. These techniques are effective in model compression to make a small model and preserve high performance. However, most model compression strategies additionally require post-processing or fine-tuning procedures to recover the lost classification performance \cite{han2015deep, wu2016quantized}, while KD does not necessarily require any further training processes for fine-tuning. Also, KD has shown great performance and been widely used to build a real-time system \cite{thai2022real, angarano2023generative, remigereau2022knowledge}. To make a robust and efficient model, we adopt KD with using knowledge from a larger and more complex model.

\subsection{Knowledge Distillation}

Knowledge distillation is one of promising techniques to train a small model in supervision of a large model. KD was firstly explored by Buciluǎ \emph{et al.} \cite{bucilua2006model} and more developed by Hinton \emph{et al.} \cite{hinton2015distilling}. Soft labels having richer information than hard labels (labeled data), outputs of a teacher network, are used in KD. Soft label enables a student network to easily mimic the softened class scores of the teacher trained with hard labels alone. For traditional KD, a student is trained with the loss function as follows:
\begin{equation}
    \mathcal{L} = (1- \lambda)\mathcal{L_{CE}} + \lambda \mathcal{L_{KD}},
\end{equation}
where, $\mathcal{L_{CE}}$ is the standard cross entropy loss, $\mathcal{L_{KD}}$ is KD loss, and $\lambda$ is a hyperparameter; $0 < \lambda < 1$.
The error between the output of the softmax layer for a student network and the ground-truth label is penalized by the cross entropy loss:
\begin{equation}
    \mathcal{L_{CE}} = \mathcal{Q}(\sigma(l_{S}), y),
\end{equation}
where, $\mathcal{Q(\cdot)}$ is a cross entropy loss function, $\sigma(\cdot)$ is a softmax function, $l_S$ is the logits of a student, and $y$ is a ground truth label. 
The outputs of student and teacher are matched by KL-divergence loss:
\begin{equation}\label{eq3}
    \mathcal{L_{KD}} = \tau^{2}KL(p_{T}, p_{S}),
\end{equation}
where, $p_T = \sigma(l_T/\tau)$ is a softened output of a teacher network, $p_S = \sigma(l_S/\tau)$ is a softened output of a student, and $\tau$ is a hyperparameter; $\tau > 1$.
The standard KD is to use a fully trained teacher and student networks. Recent studies show the effectiveness of early stopping for KD (ESKD), which utilizes early stopped model of a teacher to produce a better student than the standard knowledge distillation (Full KD) \cite{cho2019efficacy}. For the best performance, ESKD is adopted to this paper, improving the efficacy of KD \cite{cho2019efficacy}.

To transfer more effective knowledge from a teacher network, feature-based distillation using intermediate layers has been proposed \cite{gou2021knowledge, zagoruyko2016paying, tung2019similarity}. Zagoruyko \emph{et al.} \cite{zagoruyko2016paying} suggest attention transfer (AT), which uses intermediate layers to extract a map by a sum of squared attention mapping function. Tung \emph{et al.} \cite{tung2019similarity} utilizes similarity between a mini-batch of samples from a teacher, which must be matched to those from a student. The activation maps of the teacher and student have the same dimension size, which is determined by size of the mini-batch. In details, the activation map $G' \in \mathbb{R}^{b \times b}$ is produced as follows:
\begin{equation}\label{eq4}
    G' = A \cdot A^{\top} ;  A \in \mathbb{R}^{b \times chw},
\end{equation}
where, $A$ is reshaped features from an intermediate layer of a model, $b$ is the size of a mini-batch, $c$ is the number of output channels, and $h$ and $w$ are the height and width of the output, respectively. These methods are popularly used to improve the performance, however, they generally deal with uni-modal problems with a single teacher.
On the other hand, using of multiple teachers to transfer more information has been investigated \cite{gou2021knowledge, liu2020adaptive, zhang2022confidence}. Multiple teachers can provide more useful knowledge to generate a better student. Since different teachers can provide diverse knowledge, more richer information can be transferred to a student. Knowledge from teachers can be utilized individually or integrated to train a student. However, a data sample or label utilized for training a teacher cannot always be used to train/test a student \cite{gou2021knowledge}. Also, leveraging different modalities in KD increases the knowledge gap between a teacher and student, which is a factor in performance degradation \cite{gou2021knowledge}.
To resolve the problem and capture the superior knowledge, we develop a framework in KD to use topological features and two teachers for providing richer information and training a student model that does not use PIs from TDA as an input.
%Different from existing knowledge distillation methods, we 000,
The details of the proposed method is explained in section \ref{sec:proposed_method}.

\subsection{Simulated Annealing}

Simulated annealing was first introduced by Kirkpatrick \emph{et al.} \cite{kirkpatrick1983optimization} and has been used to solve optimization problems in various applications \cite{yang2020nature}. Recently, it was applied to solve KD related problems. Born-again multitask network (BAM) \cite{clark2019bam} uses a few single-task teachers to generate a multi-task student. A dynamic weighted loss for the outputs of a teacher and ground truth are used to train a student. In the early epochs of training, the student model is mostly trained by the teacher, but later, it is mostly trained by hard labels. Annealing KD \cite{jafari2021annealing} presented two stages to reduce the capacity gap between the outputs of a teacher and student. In the first stage, a temperature of KD decreases as the epoch grows while the logits of a teacher and student are matched in a regression task. In the second stage, the student is fine-tuned with hard labels by cross entropy loss.
Different from existing annealing methods \cite{clark2019bam, jafari2021annealing}, we propose a strategy of using two teachers with KD to facilitate fast saturation and reduce the knowledge gap. For the proposed method, two teachers are trained with different types of data -- time-series and persistence image data -- and their student is trained with the raw time-series data only. So, the statistical features of two teachers are different, and their distillation effects on a student are not the same. To consider the different properties of teachers and the student in distillation, we apply an annealing strategy in KD, which reduces the search space for fast saturation and helps to mitigate a knowledge gap issue by leveraging the weights of a model trained from scratch. Our method is described in the next section.

\section{Proposed Method} \label{sec:proposed_method}

For the proposed method, two teachers learned with different data are used to train a student. Firstly, to leverage topological features, we extract PIs from PDs of time-series data using TDA. Two teacher models are trained with time-series data and extracted PIs, respectively. Secondly, orthogonal features from fused correlation maps of teachers are used for distillation, considering differently activated features from teachers. In the third step, we apply an annealing strategy for knowledge distillation to optimize the weight of the student model, taking into account the time-series properties inherent in the model. Finally, a student model preserving topological features is distilled. The details of the proposed method are explained in the following section.

\subsection{Extracting Persistence Image}

Topological features provide complementary information to improve the performance in machine learning \cite{gholizadeh2018short, zeng2021topological, som2020pi}. To leverage topological features, we first extract PIs to train a model. We use Scikit-TDA python library \cite{scikittda2019} and the Ripser package for producing PDs, referring to a previous study \cite{som2020pi}. PDs of level-set filtration for time-series signals are calculated by the library. Scalar field topology presents a summary for different peaks in the signal. The PD for each channel of a sample is computed. And then, PIs are extracted from PDs based on the birth-time vs. lifetime information. We set the matrix size of the PIs as $b \times b$. The dimension size of one PI is $b\times b \times c$, where, $c$ is the number of channels for a sample. Secondly, we train a model on the extracted PIs in supervised learning. %The model is used to provide topological features to improve the classification performance. We use the pre-trained model as a teacher in KD to utilize topological features.
The model is used as the pre-trained model as a teacher, transferring topological features to a student model.

\subsection{KD with Multiple Teachers}

In test-time, generating PIs requires a large computational burden. To this end, we adopt KD to distill a student model using only time-series data as an input and learning topological features from a teacher.

\subsubsection{Distillation with Logits of Different Teachers} 
%Knowledge from logits of teachers trained with time-series data and PIs are transferred separately without using integration.
For the proposed method, since the knowledge from two teachers is transferred separately, additional processing for concatenation and hidden layers is not necessarily required.
To utilize features from two teachers, KD loss can be written as: 
\begin{equation}
    \mathcal{L_{KD}}_m = \tau^{2} \left( \alpha KL(p_{T_1}, p_{S}) + (1-\alpha) KL(p_{T_2}, p_{S})\right),
\end{equation}
where, $\alpha$ is a hyperparameter to balance the losses from different teachers, and $p_{T_1}$ and $p_{T_2}$ are softened outputs of teachers trained with time-series data and PIs, respectively.

\subsubsection{Similarity of Different Teachers} 

For better distillation, we use features from intermediate layers of teachers. However, the architectures of the teachers and student are different, and their data used for training are also different modalities. Using methods similar those proposed in Tung \emph{et al.} \cite{tung2019similarity}, we extract activation similarity matrices $G' \in \mathbb{R}^{b \times b}$ to use activated features with the same dimension size from the two teachers and student, as explained in equation \ref{eq4}.
The pattern of the activation map is highly related to the same or a different class. In details, two inputs in the same category generate the similar activation maps from a teacher, which is a beneficial to guide a student to acquire the knowledge of the teacher. % produce similar map 

\begin{figure}[htb!]
\centering
\includegraphics[width=3.0in]{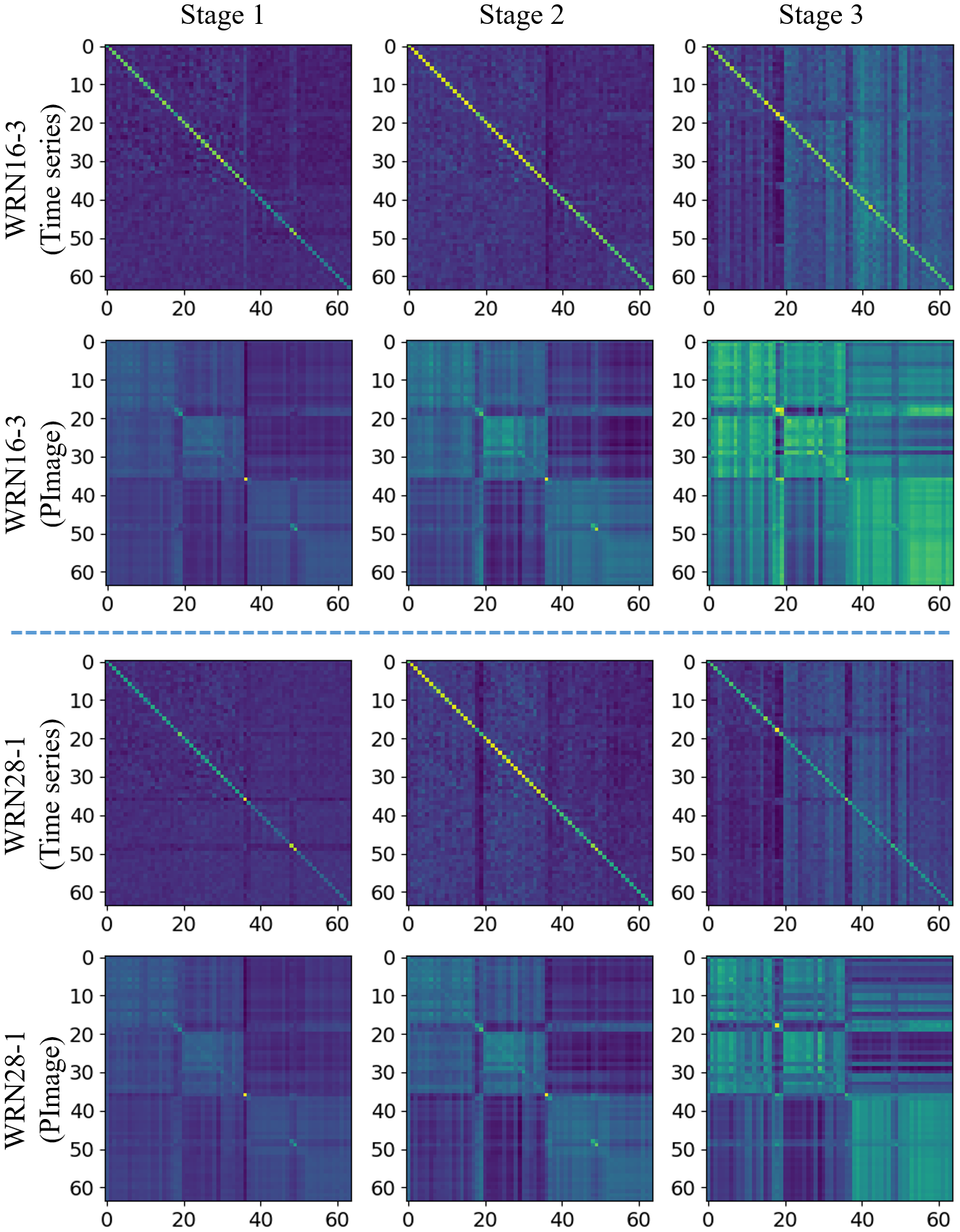}
\caption{Examples of activation similarity maps $G'$ produced by a layer for the indicated stage of the network for a batch on GENEActiv. High similarities for samples of the batch are represented with high values. The blockwise pattern is more distinctive for WRN16-3 networks, implying the higher capacity of this network can well capture the semantics of the dataset.}
\label{figure:sp_sc_map}
\end{figure}

However, since of the information gap from different modalities, difficulties still exist to transferring the each different knowledge \cite{gou2021knowledge}. As shown in Figure \ref{figure:sp_sc_map}, two models trained with different data generate dissimilar activations. These differences from multimodality make difficulties in interpreting and fusing the content, which may mislead the student \cite{kwon2020adaptive, gou2021knowledge}. To solve this issue, we create a map from two teachers by merging the activation maps with the weight value of $\alpha$ parameter as follows:
\begin{equation}
    G^{(l)}_T = \alpha G'^{(l^{T_1})}_{T_1} + (1-\alpha) G'^{(l^{T_2})}_{T_2},
\end{equation}
where, $G^{(l)}_T \in \mathbb{R}^{b \times b}$ is the generated map from the activation maps of a layer pair ($l^{T_1}$ and  $l^{T_2}$) of two teachers $G'_{T_1}$ and $G'_{T_2}$. By merging the maps, the similarities between two teachers are more highlighted. 

%layer pairs ($l$ and $l^\prime$)

%And, based on features from different teachers, the effects from  noisy feature from one teacher can be mitigated by the other teacher in transferring the knowledge. 

\subsubsection{Extracting and Transferring Orthogonal Features}

\begin{figure}[]
\centering
\includegraphics[width=3.5in]{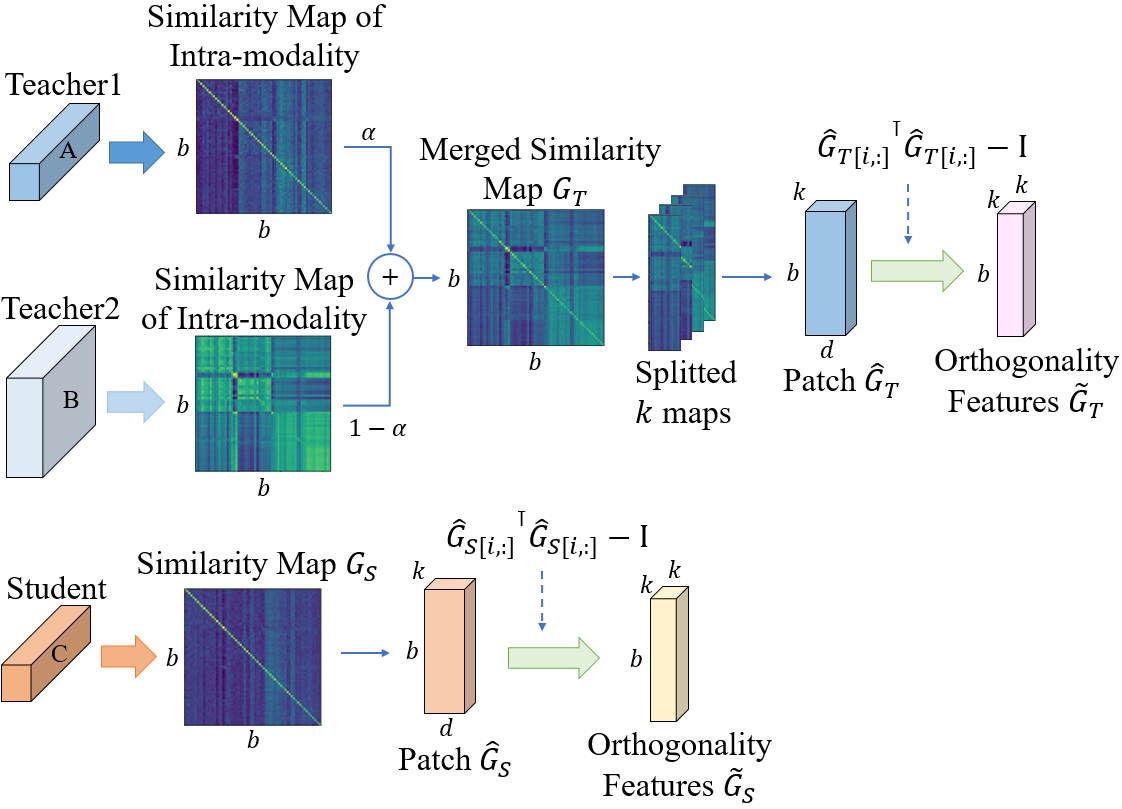}
\caption{Framework of extracting orthogonal features. A and B denote mini-batch features at a layer of Teacher1 and Teacher2, respectively. C denotes mini-batch features at a layer of Student.}
\label{figure:orthf_framework}
\end{figure}

In the ideal case, if features from multiple teachers are correlated, the errors of one teacher would not essentially affect the other one \cite{park2020orthogonality}. Since features from different teachers are merged and the data used for training a student is different from that of the teachers, it is difficult to guarantee that the teachers and student are well correlated, so the merged map from teachers may not always be good for distillation. In previous studies \cite{wang2020orthogonal, choi2020role}, orthogonality properties improve better feature explanation and lead to provide various desirable features, which enables a model to easily learn more diverse and expressive features. %Given the insight, to capture the better quality of knowledge, we construct new patches representing orthogonal features. 
Given the insight, to capture the better explanative information accounting for modality gap, we design new knowledge reflecting orthogonal properties by transforming the merged map into several patches to produce more attentive feature relationship. %to represent the more attentive feature relationship.
The overview of extracting orthogonal features is described in Figure \ref{figure:orthf_framework}.
An input-patch-matrix $\widehat{G} \in \mathbb{R}^{bd \times k}$ can be constructed by unrolling the $G$/$\norm{G}_{2}$, the normalized $G$, into $k$ columns of the matrix, where, $k$ is the number of partitions and $d$ is the size of each partition for $b$. By using the computed patch-matrix, new knowledge encoding feature relationships based on orthogonal properties is defined as follows:
%the off-diagonal elements \es{goes to zero} as:
\begin{equation}
    \widetilde{G}_{[i,:]} = \widehat{G}_{[i,:]}^{\top}\widehat{G}_{[i,:]} - \mathbf{I} ,
\end{equation}
where, $\widetilde{G}_{[i,:]} \in \mathbb{R}^{k \times k}$ represents $k\times k$ knowledge patches for $i$th element of $b$, involving orthogonality properties, and $\mathbf{I} \in k\times k$ is identity matrices. From the merged map $G_T$ and a map for the student $G_S$, $\widetilde{G_T}$ and $\widetilde{G_S}$ can be generated, respectively. Finally, the knowledge reflecting feature relationships from teachers are transferred to the student by minimizing the difference between two maps of each corresponding layer:
\begin{equation}
    \mathcal{L}_{Oth} = \frac{1}{|L|}
    \sum_{(l, l^S) \in L} \bignorm{ \widetilde{G^{(l)}_{T}} - \widetilde{G^{(l^S)}_{S}} }^{2}_{F},
\end{equation}
where, $L$ collects the layer pairs ($l$ and $l^S$), and $\norm{\cdot}_F$ is the Frobenius norm \cite{tung2019similarity}. In this way, the student is encouraged to get the similar features to the merged teacher. Therefore, the student can preserve topological as well as time-series features, which uses the raw time-series data only as an input.
%enforces % correlated/decorrelated features
The overall learning objective of the proposed method can be written as: 
\begin{equation}
    \mathcal{L}_{TP} = (1- \lambda)\mathcal{L_{CE}} + \lambda \mathcal{L_{KD}}_m + \beta\mathcal{L}_{Oth},
\end{equation}
where, $\beta$ is a hyperparameter to control the effect of loss $\mathcal{L}_{Oth}$.

%
%Even if the merged map represents more attentive features than single maps, there is a case that features from a teacher boost uncorrelated parts, off-diagonal elements, of the other teacher and some de-correlated parts remain with high values,
%which are the one of factors for degradation in distillation. To alleviate noisy effects, we extract orthogonal features from the map.

\subsection{Annealing Strategy for Multiple Teachers}
Since teachers and student are trained with different data, the models develop different statistical properties in their internal representations. Their architectures are even different, which produces more statistical gap between features from models and difficulties in training a student \cite{gou2021knowledge, jafari2021annealing}. To reduce the effects of this knowledge gap, we apply an annealing strategy in KD for the proposed method. Before training a student, we train a small model from scratch with time-series data, where the model has the same architecture as the student. When weight values are initialized to train the student, the values are determined by the pre-trained model, instead of randomly chosen values. In this way, the knowledge gap between teachers and student is mitigated and the search space for optimization is reduced. Also, this initialization enforces the student to get features that can perform well with time-series data while teachers provide their own features.

\section{Experiments} \label{sec:experiments}
In this section, we describe datasets used for evaluation and experimental settings. We evaluate the proposed method with various teacher-student combinations on wearable sensor data. We investigate the sensitivity of the proposed distillation with various hyperparameters ($\alpha$, $\beta$, and $k$). And, we explore the effectiveness of TPKD with visualization of feature maps, feature similarity analysis, and generalizability analysis. Also, we measure computational time with different methods.

\subsection{Data Description and Experimental Settings}
%\es{A. Data Description and Experimental Settings}

\subsubsection{Data Description}
We evaluate the proposed method with wearable sensor data on GENEActiv and PAMAP2 datasets.

\begin{figure}[htb!] %t!
%\begin{tabular}
\includegraphics[width = 0.6\textwidth]
{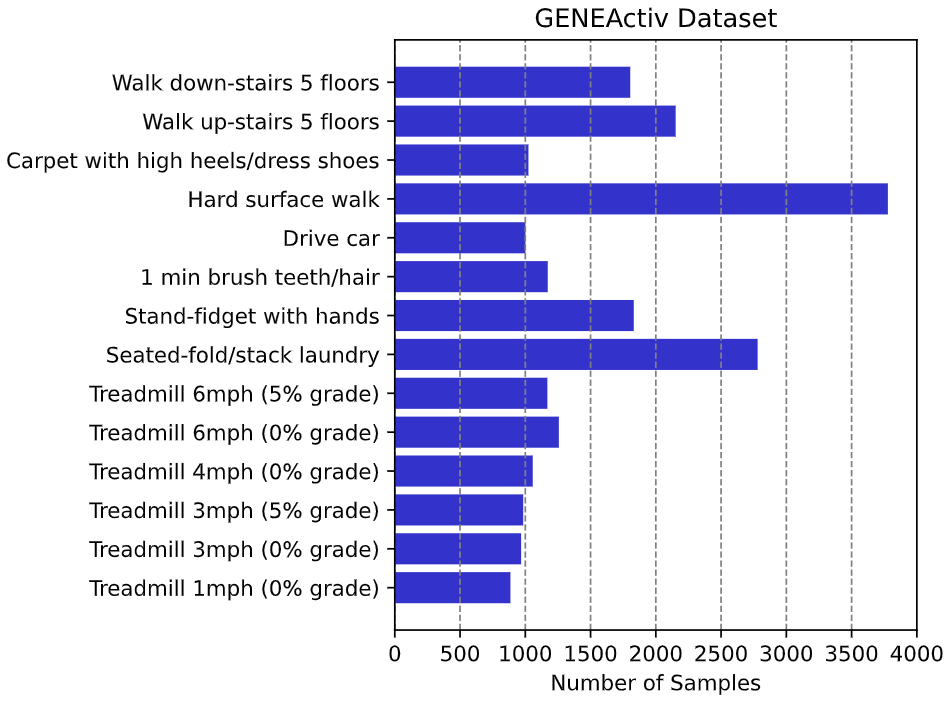} %0.27
\centering
\caption{Details of GENEActiv dataset. Each sample has 500 time-steps. From bottom to top, labels correspond to classes 0 to 13.}
%\end{tabular}
\label{figure:GENEActiv}
\end{figure}

\textbf{GENEActiv.} GENEActiv \cite{wang2016statistical} is wearable sensor based activity dataset, collected with GENEActiv sensor which is a light-weight, waterproof, and wrist-worn tri-axial accelerometer with sampling frequency of 100 Hz. In this experiment, referring to the previous study \cite{jeon2022kd}, we use 14 daily activities such as walking, sitting, and standing. As described in Figure \ref{figure:GENEActiv}, the dataset has imbalanced distribution. Each class has over 900 data samples. The number of subjects for training and testing are over 130 and 43, respectively. We use full-non-overlapping window size of 500 time-steps (5 seconds) data. The number of samples for training and testing are approximately 16k and 6k, respectively.

%%%%%%%%%%%%%%%%%%%%%%%%%%%%

%%%%%%%%%%%%%%%%%%%%%%%%%%%%%%%%%%%%%

\textbf{PAMAP2.} PAMAP2 dataset \cite{reiss2012introducing} consists of 18 physical activities (12 daily and 6 optional activities) for 9 subjects, obtained by measurements of heart rate, temperature, accelerometers, gyroscopes, and magnetometers with 100Hz of sampling frequency. The sensors were placed on  hands, chest, and ankles of the subject. In experiments on this dataset, we use 12 daily activities with 40 channels recorded from the heart rate and 4 IMUs, where activities are lying, sitting, standing, walking, etc. To compare with previous methods, the recordings are downsampled to 33.3Hz. We evaluate methods with leave-one-subject-out combination. There is missing data for some subjects and the dataset has non-uniform distribution. We use 100 time-steps (3 seconds) of a sliding window for a sample and 22 time-steps or 660 ms of step size for segmenting the sequences, which allows semi-non-overlapping sliding windows with 78\% overlapping \cite{reiss2012introducing}.

\subsubsection{Experimental Settings}
In extracting PIs, for GENEActiv, the parameter for the Gaussian function in PD is 0.25 and the values for birth-time range of PI are set as $[$-10, 10$]$, as do the previous study \cite{som2020pi}. For PAMAP2, Gaussian parameter and the birth-time range are 0.015 and $[$-1, 1$]$, respectively. Each calculated PI is normalized by its maximum value. To train network models, we set the total epochs as 200 using SGD with momentum of 0.9, the batch size as 64, and a weight decay as $1 \times 10^{−4}$. To train a model with time-series data on both datasets, the initial learning rate $lr$ is 0.05 which decreases by 0.2 at 10 epochs and drops down by 0.1 every [${t \over 3}$] where, $t$ is the total number of epochs. For training a model with image data on GENEActiv, the initial learning rate $lr$ is set to 0.1 and decreases by 0.5 at 10 epochs and drops down by 0.2 at 40, 80, 120, and 160 epochs. For PAMAP2 with image data, the initial learning rate $lr$ is set as 0.1 that drops down by 0.2 at 40, 80, 120, and 160 epochs.
For constructing teacher and student models, we use WideResNet (WRN) \cite{zagoruyko2016wide} to evaluate the performance of the proposed method, which is popularly used to validate in KD \cite{cho2019efficacy, jeon2022kd}. The model for training with time-series data consists of 1D convolutional layers, on the other hand, the one with image data consists of 2D convolutional layers. We determine $\tau$ and $\lambda$ for GENEActiv as 4 and 0.7, and for PAMAP2 as 4 and 0.99, respectively, as the previous
works do \cite{jeon2022kd}. To obtain the best results, we set optimal $\alpha$ as 0.7 for GENEActiv and 0.3 for PAMAP2, respectively. We run 3 times and report with the best averaged accuracy and standard deviation for the following experiments.
We perform baseline comparisons with traditional KD \cite{hinton2015distilling}, attention transfer (AT) \cite{zagoruyko2016paying}, similarity-preserving knowledge distillation (SP) \cite{tung2019similarity}, and simple knowledge distillation (SimKD) \cite{chen2022knowledge}, which are popularly used for distillation. $\alpha_{AT}$ and $\gamma_{SP}$ are set as 1500 and 1000 for GENEActiv, and 3500 and 700 for PAMAP2, respectively. Additionally, we compare with DIST \cite{huang2022knowledge}, which considers intra- and inter-class relationship for knowledge transfer.
Also, we compare with multi-teacher based approaches such as AVER \cite{you2017learning}, EBKD \cite{kwon2020adaptive}, CA-MKD \cite{zhang2022confidence} and AdTemp \cite{ejasilomar}. Since we use different dimensional input data and structured teachers, only the outputs from the last layer (logits) are used for baselines in distillation. % to utilize multiple teachers. %\es{explain parameter of baselines}

\subsection{Various Capacity of Teachers} \label{various_capacity}

In this section, we explore the proposed method with various capacity of teachers which are trained with time-series data and PIs, respectively. Details of models for teachers and a student, used for experiments, are summarized in Table \ref{table:info_settings}, representing model complexity and the number of trainable parameters.

%%%%%%%%%%%%%%%%%%%%%%%%%%%%%%5

\begin{table*}[htb!]
\centering
\caption{Details of teacher and student network architectures. Compression ratio is calculated with two teachers.}

\begin{center}
\scalebox{0.78}{
\begin{tabular}{c |c |c |c |c |c |c |c| c| c}
\hline
\centering
\multirow{2}{*}{DB}& Teacher1 (1D CNNs) \& & \multirow{2}{*}{Student} & \multicolumn{3}{c|}{FLOPs} & \multicolumn{3}{c|}{\# of params} & Compression  \\  \cline{4-9}
& Teacher2 (2D CNNs) & & Teacher1 & Teacher2 & Student & Teacher1 & Teacher2 & Student & ratio  \\
 \hline %\hline
\multirow{4}{*}{\rotatebox[origin=c]{90}{\scriptsize{GENEActiv}}}& WRN16-1 & \multirow{4}{*}{WRN16-1} & 11.03M & 108.97M & \multirow{4}{*}{11.03M} & 0.06M & 0.18M & \multirow{4}{*}{0.06M} & 25.93$\%$ \\ 
& WRN16-3 &  & 93.95M & 898.52M &  & 0.54M & 1.55M & & 2.94$\%$ \\ 
& WRN28-1 &  & 22.22M & 224.28M & & 0.13M & 0.37M & & 12.36$\%$ \\ 
& WRN28-3 & & 192.01M & 1923.93M & & 1.12M & 3.29M & & 1.39$\%$ \\ 
\hline

\multirow{4}{*}{\rotatebox[origin=c]{90}{\scriptsize{PAMAP2}}}& WRN16-1 & \multirow{4}{*}{WRN16-1} & 2.39M & 131.02M & \multirow{4}{*}{2.39M} & 0.06M & 0.18M & \multirow{4}{*}{0.06M} & 25.88$\%$ \\ 
& WRN16-3 & & 19.00M & 921.03M &  & 0.54M & 1.56M & & 3.01$\%$ \\ 
& WRN28-1 & & 4.64M & 246.56M & & 0.13M & 0.37M & & 12.52$\%$ \\ 
& WRN28-3 & & 38.64M & 1947.13M & &  1.12M & 3.30M & & 1.43$\%$ \\ 
\hline

\end{tabular}
}
\end{center}

\label{table:info_settings}
\end{table*}

%%%%%%%%%%%%%%%%%%%%%%%%%%%%%%%%%%%%
\begin{table}[htb!]
\centering
\renewcommand{\tabcolsep}{1.0mm} 
\caption{Accuracy ($\%$) with various knowledge distillation methods for different capacity of teachers on GENEActiv.}

\begin{center}
\scalebox{0.69}{
\begin{tabular}{c| c |c c| c c }

\hline

\multicolumn{2}{c|}{Teacher1} & WRN16-1 & WRN16-3 & WRN28-1 & WRN28-3 \\
\multicolumn{2}{c|}{(1D CNNs)} & (67.66)  & (68.89) & (68.63) & (69.23) \\

\hline
\multicolumn{2}{c|}{Teacher2} & WRN16-1 & WRN16-3 & WRN28-1 & WRN28-3 \\
\multicolumn{2}{c|}{(2D CNNs)} & (58.64) & (59.80) & (59.45) & (59.69)  \\

\hline
\multicolumn{2}{c|}{Student} & \multicolumn{4}{c}{WRN16-1}  \\ 
\multicolumn{2}{c|}{(1D CNNs)} & \multicolumn{4}{c}{(67.66{\scriptsize$\pm$0.45})} \\
\hline %\cite{hinton2015distilling}

\multirow{2}{*}{\rotatebox[origin=c]{90}{\scriptsize{PI}} }

%\multicolumn{2}{c|}{\multirow{2}{*}{Time series}} & 69.71 & 69.50 & 67.59 & 68.01\\
% \multicolumn{2}{c|}{} & {\scriptsize$\pm$0.38} & {\scriptsize$\pm$0.10} & {\scriptsize$\pm$0.36} & {\scriptsize$\pm$0.67} \\

% \cite{zagoruyko2016paying}

%\multicolumn{2}{c|}{\multirow{2}{*}{PImage}} & 67.83 & 68.76 & 68.51 & 68.46\\
%\multicolumn{2}{c|}{} & {\scriptsize$\pm$0.17} & {\scriptsize$\pm$0.73} & {\scriptsize$\pm$0.01} & {\scriptsize$\pm$0.28}\\

& \multirow{2}{*}{KD} & 67.83 & 68.76 & 68.51 & 68.46 \\
& & {\scriptsize$\pm$0.17} & {\scriptsize$\pm$0.73} & {\scriptsize$\pm$0.01} & {\scriptsize$\pm$0.28}\\
% \cite{tung2019similarity}
\hline

\multirow{10}{*}{\rotatebox[origin=c]{90}{\scriptsize{Time-series}} }
& \multirow{2}{*}{KD} & 69.71 & 69.50 & 68.32 & 68.58\\
& & {\scriptsize$\pm$0.38} & {\scriptsize$\pm$0.10} & {\scriptsize$\pm$0.63} & {\scriptsize$\pm$0.66} \\
 
 & \multirow{2}{*}{AT} & 68.21 & 69.79 & 68.09 & 67.73\\
 & & {\scriptsize$\pm$0.64} & {\scriptsize$\pm$0.36} & {\scriptsize$\pm$0.24} & {\scriptsize$\pm$0.27} \\

 & \multirow{2}{*}{SP} & 67.20 & 67.85 & 68.71 & 67.39\\
 & & {\scriptsize$\pm$0.36} & {\scriptsize$\pm$0.24} & {\scriptsize$\pm$0.46} & {\scriptsize$\pm$0.49} \\
 
  & \multirow{2}{*}{SimKD} & 69.39 & 69.89 & 68.92 & 68.80\\
 & & {\scriptsize$\pm$0.18} & {\scriptsize$\pm$0.11} & {\scriptsize$\pm$0.40} & {\scriptsize$\pm$0.38} \\
 
  & \multirow{2}{*}{DIST} & 68.20 & 69.71 & 69.23 & 68.18\\
 & & {\scriptsize$\pm$0.28} & {\scriptsize$\pm$0.15} & {\scriptsize$\pm$0.19} & {\scriptsize$\pm$0.60} \\

\hline

%\multirow{14}{*}{\rotatebox[origin=c]{90}{\scriptsize{TS+PImage}} } & %\multirow{2}{*}{KD} & 69.09 & 69.24 & 69.55 & 69.42 \\
%&  & {\scriptsize$\pm$0.37} & {\scriptsize$\pm$0.62} & {\scriptsize$\pm$0.41} & {\scriptsize$\pm$0.58} \\
 
\multirow{16}{*}{\rotatebox[origin=c]{90}{\scriptsize{TS+PImage}} }

 & \multirow{2}{*}{AVER} & 68.99 & 68.74 & 68.77 & 69.02\\
 & & {\scriptsize$\pm$0.76} & {\scriptsize$\pm$0.35} & {\scriptsize$\pm$0.70} & {\scriptsize$\pm$0.50} \\
 
 & \multirow{2}{*}{EBKD} & 68.43 & 69.24 & 68.45 & 67.50\\
 & & {\scriptsize$\pm$0.25} & {\scriptsize$\pm$0.25} & {\scriptsize$\pm$0.73} & {\scriptsize$\pm$0.40} \\
 
 & \multirow{2}{*}{CA-MKD} & 69.33 & 69.80 & 69.61 & 68.81\\
 & & {\scriptsize$\pm$0.61} & {\scriptsize$\pm$0.16} & {\scriptsize$\pm$0.57} & {\scriptsize$\pm$0.79} \\
 
 \cline{2-6}
 
 & \multirow{2}{*}{Base} & 69.09 & 69.24 & 69.55 & 69.42 \\
 &  & {\scriptsize$\pm$0.37} & {\scriptsize$\pm$0.62} & {\scriptsize$\pm$0.41} & {\scriptsize$\pm$0.58} \\

 & \multirow{2}{*}{AdTemp} & 69.80 & 70.10 & 70.01 & 69.55\\
 & & {\scriptsize$\pm$0.68} & {\scriptsize$\pm$0.39} & {\scriptsize$\pm$0.83} & {\scriptsize$\pm$0.51} \\

 & \multirow{2}{*}{Ann.} & 70.15 & 70.71 & 70.44 & 69.97\\
 & & {\scriptsize$\pm$0.03} & {\scriptsize$\pm$0.12} & {\scriptsize$\pm$0.10} & {\scriptsize$\pm$0.06} \\
 
 & TPKD & 70.71 & 70.93 & 70.71 & 70.12 \\
 & (w/o Orth.) & {\scriptsize$\pm$0.20} & {\scriptsize$\pm$0.26} & {\scriptsize$\pm$0.14} & {\scriptsize$\pm$0.21} \\

 &  TPKD & \textbf{71.05} & \textbf{71.10} & \textbf{70.97} & \textbf{70.50} \\
 & (w/ Orth.) & {\scriptsize$\pm$0.13} & {\scriptsize$\pm$0.11} & {\scriptsize$\pm$0.12} & {\scriptsize$\pm$0.15} \\
\hline

\end{tabular}
}
\end{center}
%\vspace{-1.3em}
\label{table:combT_GENE}
\end{table}

%%%%%%%%%%%%%%%%%%%%%%%%%%%%%%%%%%%%

\begin{figure}[htb!] %t!
%\begin{tabular}
\includegraphics[width = 0.99\textwidth]
{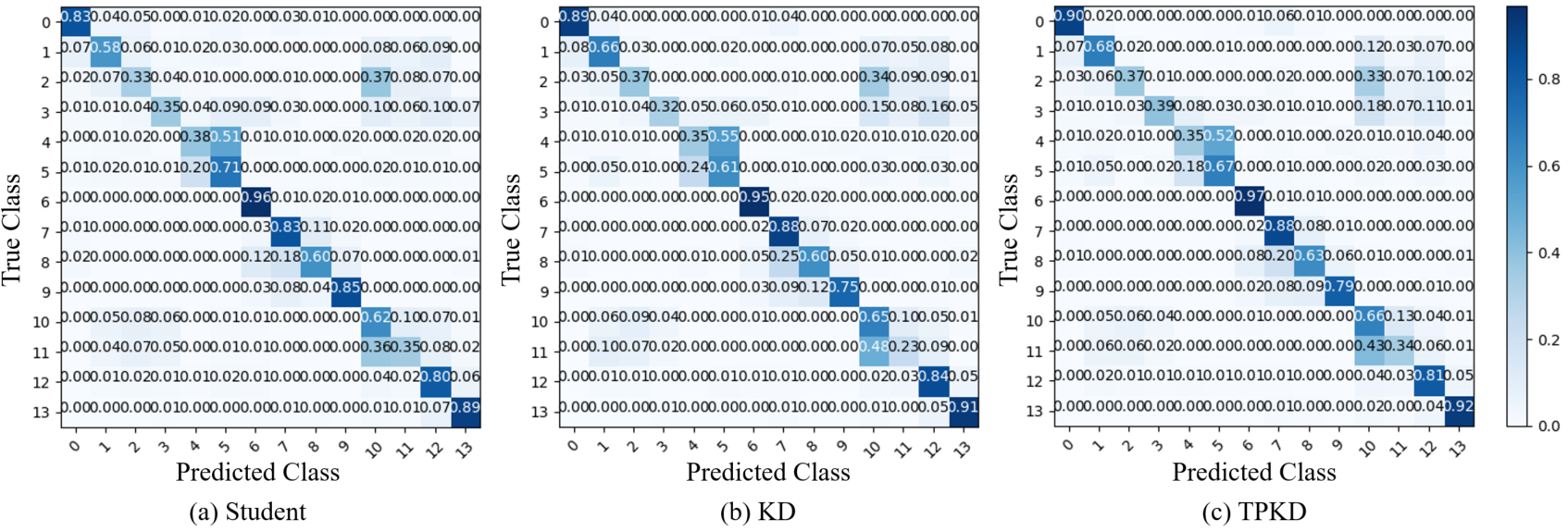} %0.27
\centering
\caption{Comparison of confusion matrices with various methods on GENEActiv.}
%\end{tabular}
\label{figure:confmatrix}
\end{figure}

%%%%%%%%%%%%%%%%%%%%%%%%%%%%%%%%%%%%%

\begin{table}[htb!]
\caption{Accuracy ($\%$) for related methods on GENEActiv with 7 classes.}\label{table:GENE_Acc_1}
\centering
\scalebox{0.75}{
\begin{tabular}{c | p{12em} |c c}
\hline
%\centering
\multicolumn{2}{c|}{\multirow{2}{*}{Method}} & \multicolumn{2}{c}{Window length} \\
%\cline{2-3}
   \multicolumn{2}{c|}{}  & 1000 & 500 \\ \hline

    \multirow{17}{*}{\rotatebox[origin=c]{90}{\scriptsize{Time-series}} }
    
    & SVM \cite{cortes1995support} & 86.29 & 85.86  \\
    & Choi \textit{et al.} \cite{choi2018temporal} & 89.43 & 87.86 \\ \cline{2-4}
    
   & WRN16-1 & 89.29{\scriptsize$\pm$0.32} & 86.83{\scriptsize$\pm$0.15} \\
    & WRN16-3 & 89.53{\scriptsize$\pm$0.15} & 87.95{\scriptsize$\pm$0.25} \\
    & WRN16-8 & 89.31{\scriptsize$\pm$0.21} & 87.29{\scriptsize$\pm$0.17} \\      \cline{2-4}
   & ESKD (WRN16-3) & 89.88{\scriptsize$\pm$0.07} & 88.16{\scriptsize$\pm$0.15} \\
   & ESKD (WRN16-8) & 89.58{\scriptsize$\pm$0.13} & 87.47{\scriptsize$\pm$0.11}  \\
    & Full KD (WRN16-3) & 89.84{\scriptsize$\pm$0.21} & 87.05{\scriptsize$\pm$0.19} \\
    & Full KD (WRN16-8) & 89.36{\scriptsize$\pm$0.06} & 86.38{\scriptsize$\pm$0.06}  \\
     %\hline
     \cline{2-4}

    & AT (WRN16-1) & 90.10{\scriptsize$\pm$0.49} & 87.25{\scriptsize$\pm$0.22}  \\
    & AT (WRN16-3) & 90.32{\scriptsize$\pm$0.09} & 87.60{\scriptsize$\pm$0.22}  \\
    & SP (WRN16-1) & 87.08{\scriptsize$\pm$0.56} & 87.65{\scriptsize$\pm$0.11}  \\
    & SP (WRN16-3) & 88.47{\scriptsize$\pm$0.19} & 87.69{\scriptsize$\pm$0.18}  \\ 
    & SimKD (WRN16-1) & 90.25{\scriptsize$\pm$0.22} & 87.24{\scriptsize$\pm$0.09}  \\
    & SimKD (WRN16-3) & 90.47{\scriptsize$\pm$0.32} & 88.16{\scriptsize$\pm$0.37}  \\
    & DIST (WRN16-1) & 90.18{\scriptsize$\pm$0.31} & 87.62{\scriptsize$\pm$0.02}  \\
    & DIST (WRN16-3) & 90.20{\scriptsize$\pm$0.39} & 87.05{\scriptsize$\pm$0.31}  \\ \hline
    \multirow{10}{*}{\rotatebox[origin=c]{90}{\scriptsize{TS+PImage}} }
    & AVER (WRN16-1) & 90.01{\scriptsize$\pm$0.46} & 87.53{\scriptsize$\pm$0.16}  \\ 
    & AVER (WRN16-3) & 90.06{\scriptsize$\pm$0.33} & 87.05{\scriptsize$\pm$0.37}  \\ 
    & EBKD (WRN16-1) & 90.35{\scriptsize$\pm$0.12} & 87.51{\scriptsize$\pm$0.41}  \\ 
    & EBKD (WRN16-3) & 89.82{\scriptsize$\pm$0.14} & 87.66{\scriptsize$\pm$0.28}  \\ 
    & CA-MKD (WRN16-1) & 90.01{\scriptsize$\pm$0.28} & 87.14{\scriptsize$\pm$0.25}  \\
    & CA-MKD (WRN16-3) & 90.13{\scriptsize$\pm$0.34} & 88.04{\scriptsize$\pm$0.26}  \\ \cline{2-4}
    & Ann. (WRN16-1) & 90.44{\scriptsize$\pm$0.16} & 88.18{\scriptsize$\pm$0.12} \\ 
    & Ann. (WRN16-3) & 90.71{\scriptsize$\pm$0.15} & 88.26{\scriptsize$\pm$0.24}  \\ 
    & TPKD (w/ Orth.) (WRN16-1) & \textbf{90.93}{\scriptsize$\pm$0.11} & \textbf{88.83}{\scriptsize$\pm$0.22}  \\
    & TPKD (w/ Orth.) (WRN16-3) & 90.83{\scriptsize$\pm$0.09} & 88.60{\scriptsize$\pm$0.25}  \\ \hline
\end{tabular}
}
\end{table}

%%%%%%%%%%%%%%%%%%%%%%%%%%%%%%%%%%%%%%%%%%%%%%%
The experimental results on GENEActiv with various teachers are described in Table \ref{table:combT_GENE}. Note, ``Time-series'' and ``PImage'' denote results of the model trained by KD with Teacher1 trained with time-series data and Teacher2 trained with PIs, respectively. ``TS'', ``Base'', and ``Ann.'' denote using a teacher trained with time-series data, a model trained by two teachers in KD using logits balanced with $\alpha$, and applying annealing strategy, respectively. ``Orth.'' denotes using orthogonal features in distillation. When TPKD is implemented without orthogonal features, the merged map of teachers and the one of student are matched directly by mean squared error in distillation. 
The numbers in brackets imply trainable parameters of the model and accuracy, respectively. From the left to right combinations of teachers in the table, ($\beta$, $k$) of TPKD are defined as (900, 4), (700, 2), (700, 4), and (900, 4), respectively. TPKD (TS+PImage with Ann.+orthogonal feature distillation), as shown in the table, achieves the best performing results in all cases. %When teachers from time-series and PIs are used together in distillation, even the basic model shows higher accuracy than a model trained from scratch and KD trained with time series data alone. Furthermore,
Base models trained with the annealing strategy (Ann.) outperform the results of baselines and the basic model (Base), indicating that the strategy aids in performance improvement. Also, a larger model does not necessarily guarantee a better student, corroborating previous observations \cite{cho2019efficacy}. In more detail, a student from WRN16-1 teachers performs better than WRN28-3 teachers.
We also plot confusion matrices for WRN16-1 students learned with WRN16-3 teachers. As shown in Figure \ref{figure:confmatrix}, class 4 and 5 cases are challenges in all methods because these classes are walking on treadmills with the same speed (6 mph) and different incline levels (0\% and 5\%). For classes 10 and 11, these classes are walking with different shoes or different surfaces. Since activities are in the same walking category, these classes are more difficult to be classified correctly. In overall results, TPKD performs best. For KD trained with time-series, results for classes 5 and 11 show much degradation compared to learning from scratch (Student). In these classes, TPKD outperforms KD, which implies that topological features complement time-series features in KD and alleviate performance degradation.

%%%%%%%%%%%%%%%%%%%%%%%%%%%%%%%%

\begin{table}[htb!]
\centering
\renewcommand{\tabcolsep}{1.2mm} 
\caption{Accuracy ($\%$) with various knowledge distillation methods for different capacity of teachers on PAMAP2.}

\begin{center}
\scalebox{0.67}{
\begin{tabular}{c| c |c c| c c }

\hline

\multicolumn{2}{c|}{Teacher1} & WRN16-1 & WRN16-3 & WRN28-1 & WRN28-3 \\
\multicolumn{2}{c|}{(1D CNNs)} & (85.27)  & (85.80) & (84.81) & (84.46) \\

\hline
\multicolumn{2}{c|}{Teacher2} & WRN16-1 & WRN16-3 & WRN28-1 & WRN28-3 \\
\multicolumn{2}{c|}{(2D CNNs)} & (86.93) & (87.23) & (87.45) & (87.88)  \\

\hline
\multicolumn{2}{c|}{Student} & \multicolumn{4}{c}{WRN16-1}  \\ 
\multicolumn{2}{c|}{(1D CNNs)} & \multicolumn{4}{c}{(82.99{\scriptsize$\pm$2.50})} \\
\hline %\cite{hinton2015distilling}

\multirow{2}{*}{\rotatebox[origin=c]{90}{\scriptsize{PI}} }

%\multicolumn{2}{c|}{\multirow{2}{*}{Time series}} & 69.71 & 69.50 & 67.59 & 68.01\\
% \multicolumn{2}{c|}{} & {\scriptsize$\pm$0.38} & {\scriptsize$\pm$0.10} & {\scriptsize$\pm$0.36} & {\scriptsize$\pm$0.67} \\

% \cite{zagoruyko2016paying}

%\multicolumn{2}{c|}{\multirow{2}{*}{PImage}} & 67.83 & 68.76 & 68.51 & 68.46\\
%\multicolumn{2}{c|}{} & {\scriptsize$\pm$0.17} & {\scriptsize$\pm$0.73} & {\scriptsize$\pm$0.01} & {\scriptsize$\pm$0.28}\\

& \multirow{2}{*}{KD} & 85.04 & 86.68 & 85.08 & 85.39 \\
& & {\scriptsize$\pm$2.58} & {\scriptsize$\pm$2.19} & {\scriptsize$\pm$2.44} & {\scriptsize$\pm$2.35}\\
% \cite{tung2019similarity}
\hline

\multirow{2}{*}{\rotatebox[origin=c]{90}{\scriptsize{TS}} }

& \multirow{2}{*}{KD} & 85.96 & 86.50 & 84.92 & 86.26\\
& & {\scriptsize$\pm$2.19} & {\scriptsize$\pm$2.21} & {\scriptsize$\pm$2.45} & {\scriptsize$\pm$2.40} \\ \hline

%\multirow{14}{*}{\rotatebox[origin=c]{90}{\scriptsize{TS+PImage}} } & %\multirow{2}{*}{KD} & 69.09 & 69.24 & 69.55 & 69.42 \\
%&  & {\scriptsize$\pm$0.37} & {\scriptsize$\pm$0.62} & {\scriptsize$\pm$0.41} & {\scriptsize$\pm$0.58} \\
 
\multirow{8}{*}{\rotatebox[origin=c]{90}{\scriptsize{TS+PImage}} }

 & \multirow{2}{*}{Base} & 85.91 & 86.18 & 85.54 & 86.04 \\
 &  & {\scriptsize$\pm$2.32} & {\scriptsize$\pm$2.37} & {\scriptsize$\pm$2.26} & {\scriptsize$\pm$2.24} \\

 & \multirow{2}{*}{Ann.} & 86.09 & 87.12 & 85.89 & 86.33\\
 & & {\scriptsize$\pm$2.33} & {\scriptsize$\pm$2.26} & {\scriptsize$\pm$2.26} & {\scriptsize$\pm$2.30} \\
 
 & TPKD & 87.26 & 88.00 & 86.47 & 86.92 \\
 & (w/o Orth.) & {\scriptsize$\pm$2.09} & {\scriptsize$\pm$2.21} & {\scriptsize$\pm$2.26} & {\scriptsize$\pm$2.27} \\

 &  TPKD & \textbf{87.67} & \textbf{88.45} & \textbf{86.86} & \textbf{87.40} \\
 &  (w/ Orth.) & {\scriptsize$\pm$2.01} & {\scriptsize$\pm$2.10} & {\scriptsize$\pm$2.07} & {\scriptsize$\pm$2.13} \\
\hline

\end{tabular}
}
\end{center}
%\vspace{-1.3em}
\label{table:combT_PAMAP2}
\end{table}

\begin{table}[htb!]
\caption{Accuracy ($\%$) for related methods on PAMAP2.}
\label{table:PAMAP2_1previous}
\centering
\scalebox{0.74}{
\begin{tabular}{c |p{13em} | c } %|c c|
\hline
%\centering
\multicolumn{2}{c|}{Method} & Accuracy (\%) \\ \hline

\multirow{17}{*}{\rotatebox[origin=c]{90}{\scriptsize{Time-series}} }

& Chen and Xue \cite{chen2015deep}  &  83.06  \\
& Ha \textit{et al.}\cite{ha2015multi}  &  73.79  \\
& Ha and Choi \cite{ha2016convolutional}  &  74.21  \\
& Kwapisz \cite{kwapisz2011activity}  &  71.27  \\
& Catal \textit{et al.} \cite{catal2015use}  &  85.25  \\
& Kim \textit{et al.}\cite{kim2012analysis}  &  81.57  \\ \cline{2-3}

& WRN16-1 & 82.81{\scriptsize$\pm$2.51}  \\
& WRN16-3 & 84.18{\scriptsize$\pm$2.28}  \\
& WRN16-8 & 83.39{\scriptsize$\pm$2.26}  \\ \cline{2-3}
& ESKD (WRN16-3) & 86.38{\scriptsize$\pm$2.25} \\
& ESKD (WRN16-8) & 85.11{\scriptsize$\pm$2.46}  \\
& Full KD (WRN16-3) & 84.31{\scriptsize$\pm$2.24} \\
& Full KD (WRN16-8) & 83.70{\scriptsize$\pm$2.52}  \\ \cline{2-3}

& AT (WRN16-1) & 83.79{\scriptsize$\pm$2.40} \\
& AT (WRN16-3) & 84.44{\scriptsize$\pm$2.22}  \\
& SP (WRN16-1) & 84.31{\scriptsize$\pm$2.38} \\
& SP (WRN16-3) & 84.89{\scriptsize$\pm$2.10}  \\ \hline

\multirow{10}{*}{\rotatebox[origin=c]{90}{\scriptsize{TS+PImage}} }
& AVER (WRN16-1) & 85.82{\scriptsize$\pm$2.16} \\
& AVER (WRN16-3) & 86.00{\scriptsize$\pm$2.45}  \\
& EBKD (WRN16-1) & 85.58{\scriptsize$\pm$2.31} \\
& EBKD (WRN16-3) & 85.62{\scriptsize$\pm$2.37}  \\
& CA-MKD (WRN16-1) & 84.06{\scriptsize$\pm$2.50} \\
& CA-MKD (WRN16-3) & 85.02{\scriptsize$\pm$2.64}  \\   \cline{2-3}
& Ann. (WRN16-1) & 86.09{\scriptsize$\pm$2.33} \\
& Ann. (WRN16-3) & 87.12{\scriptsize$\pm$2.26}  \\
& TPKD (w/ Orth.) (WRN16-1) & 87.67{\scriptsize$\pm$2.01} \\
& TPKD (w/ Orth.) (WRN16-3) & \textbf{88.45}{\scriptsize$\pm$2.10}  \\  \hline

%Mean Accuracy  &  78.19  \\ \hline
\end{tabular}
}
\end{table}

%%%%%%%%%%%%%%%%%%%%%%%%%%%%%%%%%%%%%%%

To compare with different sample window lengths and more previous studies, we evaluate the methods with 7 classes of GENEActiv dataset, as do the previous study \cite{jeon2022kd, choi2018temporal}. WRN16-1 (1D CNNs) student is used. The brackets denote the teacher models. ($\beta$, $k$) of TPKD are set as (1100, 4) for WRN16-1 teachers with both window lengths and WRN16-3 teachers with window length of 500, and (500, 8) for WRN16-3 teachers with window length of 1000, respectively. As summarized in Table \ref{table:GENE_Acc_1}, results of TPKD (w/ Orth.) with WRN16-1 teachers show the best in both cases.
%When an annealing strategy is applied, smaller teachers distill better students.
Since one of teachers (WRN16-1) has the same structure of the student (WRN16-1), the knowledge gap is not much different than the larger teachers (WRN16-3). Compared to smaller length of window sizes, larger sized window samples can generate better results for all cases. This is because larger samples can provide more information that can be utilized to train models for classification task.
The results with various capacity of teachers on PAMAP2 are described in Table \ref{table:combT_PAMAP2}. $\beta$ and $k$ of TPKD are defined as 200 and 4, respectively. TPKD (w/ Orth.) shows the best in all cases. As described in Table \ref{table:PAMAP2_1previous}, TPKD outperforms the previous methods. Therefore, TPKD allows model compression and improves accuracy across datasets.

\subsection{Various Combinations of Teachers} \label{various_combt}

To understand the effect of different teacher architectures, various combinations of two teachers are used, considering different channel and depth of WRN. Results on GENEActiv and PAMAP2 are described in Table \ref{table:mixcombT_GENE} and \ref{table:mixcombT_PAMAP2}, respectively.
($\beta$, $k$) on GENEActiv for each combination is indicated in Table \ref{table:mixcombT_GENE}. ($\beta$, $k$) on PAMAP2 is set as (200, 4). $\beta$ for TPKD without using orthogonal features is set as 700 and 200 on GENEActiv and PAMAP2, respectively.

%%%%%%%%%%%%%%%%%%%%%%%%%%%%%%%%%%%%%%%%%%% various combination of teachers
\begin{table*}[htb!]
\centering
\renewcommand{\tabcolsep}{1.2mm} 
\caption{Accuracy ($\%$) with various knowledge distillation methods for different structure of teachers on GENEActiv.} 

\begin{center}
\scalebox{0.85}{
\begin{tabular}{c |c c c c| c c c c | c c c c}

\hline
\centering

\multirow{2}{*}{Method} & \multicolumn{12}{c}{Architecture Difference} \\ \cline{2-13}
 & \multicolumn{4}{c|}{Depth} &
\multicolumn{4}{c|}{Width} & \multicolumn{4}{c}{Depth+Width} \\
\hline

 & WRN & WRN & WRN & WRN &
WRN & WRN & WRN & WRN & WRN & WRN & WRN & WRN \\
Teacher1 & 16-1 & 16-1 & 28-1 & 40-1 &
16-1 & 16-3 & 28-1 & 28-3 & 28-1 & 28-3 & 40-1 & 16-1 \\
(1D CNNs) & (0.06M, & (0.06M, & (0.1M, & (0.2M, &
(0.06M, & (0.5M, & (0.1M, & (1.1M, & (0.1M, & (1.1M, & (0.2M, & (0.06M, \\
& 67.66) & 67.66) & 68.63) & 69.05) &
67.66) & 68.89) & 68.63) & 69.23) & 68.63) & 69.23) & 69.05) & 67.66)  \\ \hline

 & WRN & WRN & WRN & WRN &
WRN & WRN & WRN & WRN & WRN & WRN & WRN & WRN \\
Teacher2 & 28-1 & 40-1 & 16-1 & 16-1 &
16-3 & 16-1 & 28-3 & 28-1 & 16-3 & 40-1 & 28-3 & 28-3 \\
(2D CNNs) & (0.4M, & (0.6M, & (0.2M, & (0.2M, &
(1.6M, & (0.2M, & (3.3M, & (0.4M, & (1.6M, & (0.6M, & (3.3M, & (3.3M, \\
& 59.45) & 59.67) & 58.64) & 58.64) &
59.80) & 58.64) & 59.69) & 59.45) & 59.80) & 59.67) & 59.69) & 59.69)  \\

\hline

Student & \multicolumn{12}{c}{WRN16-1}  \\ 
(1D CNNs) & \multicolumn{12}{c}{(0.06M, 67.66{\scriptsize$\pm$0.45})} \\ \hline
 %\cite{hinton2015distilling}

\multirow{2}{*}{Base} & 68.71 & 68.41 & 67.89 & 68.33 &
68.77 & 68.92 & 68.26 & 69.09 & 68.04 & 68.29 & 68.90 & 68.15 \\
 & {\scriptsize$\pm$0.36} & {\scriptsize$\pm$0.27} & {\scriptsize$\pm$0.27} & {\scriptsize$\pm$0.17} &
{\scriptsize$\pm$0.43} & {\scriptsize$\pm$0.79} & {\scriptsize$\pm$0.13} & {\scriptsize$\pm$0.59} & {\scriptsize$\pm$0.24} & {\scriptsize$\pm$0.27} & {\scriptsize$\pm$0.50} & {\scriptsize$\pm$0.23}  \\
% \cite{zagoruyko2016paying}
\multirow{2}{*}{Ann.} & 69.95 & 69.86 & 70.34 & 70.56 &
69.68 & 71.06 & 70.28 & 69.95 & 70.28 & 69.87 & 70.49 & 69.65  \\
 & {\scriptsize$\pm$0.05} & {\scriptsize$\pm$0.07} & {\scriptsize$\pm$0.14} & {\scriptsize$\pm$0.04} &
{\scriptsize$\pm$0.14} & {\scriptsize$\pm$0.02} & {\scriptsize$\pm$0.08} & {\scriptsize$\pm$0.07} & {\scriptsize$\pm$0.13} & {\scriptsize$\pm$0.23} & {\scriptsize$\pm$0.05} & {\scriptsize$\pm$0.04}  \\
% \cite{tung2019similarity}
\multirow{2}{*}{\makecell{TPKD \\ (w/o Orth.)}} & 70.39 & 70.47 & 71.01 & 71.36 &
69.82 & 71.11 & 70.53 & 70.31 & \textbf{70.55} & 70.57 & 70.55 & 70.68 \\
 & {\scriptsize$\pm$0.12} & {\scriptsize$\pm$0.40} & {\scriptsize$\pm$0.04} & {\scriptsize$\pm$0.06} &
{\scriptsize$\pm$0.23} & {\scriptsize$\pm$0.18} & {\scriptsize$\pm$0.26} & {\scriptsize$\pm$0.15} & {\scriptsize$\pm$0.28} & {\scriptsize$\pm$0.18} & {\scriptsize$\pm$0.22} & {\scriptsize$\pm$0.10} \\

%\hline

\multirow{3}{*}{\makecell{TPKD \\ (w/ Orth.) \\ ($\beta$, $k$)}} & \textbf{70.67} & \textbf{70.76} & \textbf{71.74} & \textbf{71.40} &
\textbf{70.03} & \textbf{71.25} & \textbf{71.08} & \textbf{70.35} & 70.42 & \textbf{70.65} & \textbf{71.04} & \textbf{71.00} \\
 & {\scriptsize$\pm$0.33} & {\scriptsize$\pm$0.22} & {\scriptsize$\pm$0.07} & {\scriptsize$\pm$0.05} &
{\scriptsize$\pm$0.14} & {\scriptsize$\pm$0.18} & {\scriptsize$\pm$0.21} & {\scriptsize$\pm$0.09} & {\scriptsize$\pm$0.21} & {\scriptsize$\pm$0.24} & {\scriptsize$\pm$0.29} & {\scriptsize$\pm$0.33}  \\
& (900, 4) & (900, 4) & (700, 4) & (900, 4)
& (700, 4) & (700, 2) & (900, 4) & (700, 4) &
(1100, 4) & (900, 4) & (700, 4) & (900, 4) \\

\hline

\end{tabular}
}
\end{center}
%\vspace{-1.3em}
\label{table:mixcombT_GENE}
\end{table*}
%%%%%%%%%%%%%%%%%%%%%%%%%%%%%%%%%%%%%%%%%%%

%%%%%%%%%%%%%%%%%%%%%%%%%%%%%%%%%%%%%%%%%%%
%\es{PAMAP2}

\begin{table}[htb!]
\centering
\renewcommand{\tabcolsep}{1.7mm} 
\caption{Accuracy ($\%$) with various knowledge distillation methods for different structure of teachers on PAMAP2.}

\begin{center}
\scalebox{0.75}{
\begin{tabular}{c |c c |c | c c c}

\hline
\centering
\multirow{2}{*}{Method} & \multicolumn{6}{c}{Architecture Difference} \\ \cline{2-7}
& \multicolumn{2}{c|}{Depth} &
Width & \multicolumn{3}{c}{Depth+Width} \\
\hline

 & WRN & WRN & WRN & WRN &
WRN & WRN\\
Teacher1 & 28-1 & 16-1 & 28-3 & 16-3 &
16-1 & 28-3 \\
(1D CNNs) & (0.1M, & (0.06M, & (1.1M, & (0.5M, &
(0.06M, & (1.1M, \\
& 84.81) & 85.27) & 84.46) & 85.80) &
85.27) & 84.46) \\ \hline

 & WRN & WRN & WRN & WRN &
WRN & WRN \\
Teacher2 & 16-1 & 28-1 & 28-1 & 28-1 &
28-3 & 16-1 \\
(2D CNNs) & (0.2M, & (0.4M, & (0.4M, & (0.4M, &
(3.3M, & (0.2M, \\
& 86.93) & 87.45) & 87.45) & 87.45) &
87.88) & 86.93) \\

\hline

Student & \multicolumn{6}{c}{WRN16-1}  \\ 
(1D CNNs) & \multicolumn{6}{c}{(0.06M, 82.99{\scriptsize$\pm$2.50})} \\ \hline
 %\cite{hinton2015distilling}

% \multirow{2}{*}{Base} & 68.71 & 68.41 & 67.89 & 68.33 &
% 68.77 & 68.92 \\
%  & {\scriptsize$\pm$0.36} & {\scriptsize$\pm$0.27} & {\scriptsize$\pm$0.27} & {\scriptsize$\pm$0.17} &
% {\scriptsize$\pm$0.43} & {\scriptsize$\pm$0.79} \\

\multirow{2}{*}{Ann.} & 85.97 & 85.33 & 85.59 & 85.82 &
85.94 & 85.86 \\
 & {\scriptsize$\pm$2.33} & {\scriptsize$\pm$2.22} & {\scriptsize$\pm$2.28} & {\scriptsize$\pm$2.26} &
{\scriptsize$\pm$2.31} & {\scriptsize$\pm$2.42} \\

% \multirow{2}{*}{Ann.+SP} & 70.39 & 70.47 & 71.01 & 71.36 &
% 69.82 & 71.11 \\
%  & {\scriptsize$\pm$0.12} & {\scriptsize$\pm$0.40} & {\scriptsize$\pm$0.04} & {\scriptsize$\pm$0.06} &
% {\scriptsize$\pm$0.23} & {\scriptsize$\pm$0.18} \\

%\hline

TPKD & \textbf{86.10} & \textbf{87.26} & \textbf{87.94} & \textbf{87.82} &
\textbf{87.02} & \textbf{86.97} \\
(w/ Orth.) & {\scriptsize$\pm$2.30} & {\scriptsize$\pm$1.96} & {\scriptsize$\pm$2.08} & {\scriptsize$\pm$2.07} &
{\scriptsize$\pm$1.98} & {\scriptsize$\pm$2.26} \\

\hline

\end{tabular}
}
\end{center}
%\vspace{-1.3em}
\label{table:mixcombT_PAMAP2}
\end{table}

%%%%%%%%%%%%%%%%%%%%%%%%%%%%%%%%%%%%%%%%%%%

As shown in Table \ref{table:mixcombT_GENE}, in most cases, TPKD (w/ Orth.) shows the best performance. When the capacity of Teacher1 is high, the result gap between baselines and TPKD tends to be small, where TPKD still performs better. When both teachers are small (e.g. WRN28-1 Teacher1 and WRN16-1 Teacher2), the student by TPKD performs better than the one from the other combinations of teachers. Also, when width of teachers is the same as the student, the proposed method shows better performance than other combinations of teachers. This implies that TPKD performs better when width of teachers are similar to a student among various combinations of teachers.

As described in Table \ref{table:mixcombT_PAMAP2}, TPKD shows better performance than Ann. (applying annealing strategy only) in all cases. When Teacher1 is WRN28-3 and Teacher2 is WRN28-1, TPKD shows the best accuracy among different combinations of teachers. This result also shows when the capacity of Teacher1 is high, the result gap between baselines and TPKD tends to be small. This is because a large teacher creates more knowledge gap which makes challenges in distillation. There is some cases that baselines produce less improvement with large Teacher1, compared to using small one. Even if the performance is affected from the knowledge gap, TPKD alleviates the negative effects in distillation, which outperforms the all baselines, and even generates a better student than its teachers. Also, the results corroborate that large teachers does not always distill a better student \cite{cho2019efficacy}.

\subsection{Ablations and Sensitivity Analysis}

In this section, we investigate the effects of hyperparameters ($\alpha$, $\beta$, and $k$) on TPKD (with orthogonal features). And, feature maps from intermediate layers of trained students are visualized to better understand the performance of TPKD. Also, we analyze feature similarities and generalizability of models. Additionally, to figure out the robustness of TPKD, we explore the proposed method under noisy testing data.

%\es{D. Sensitivity and ablation study}

\subsubsection{Effect of Distillation Hyperparameters on TPKD}
%\es{-- sensitivity on alpha and sp param -- bar graph}
%%%%%%%%%%%%%%%%%%%%%%%%%%%%%%

\begin{figure}[htb!]
\centering
\includegraphics[width=3.7in]{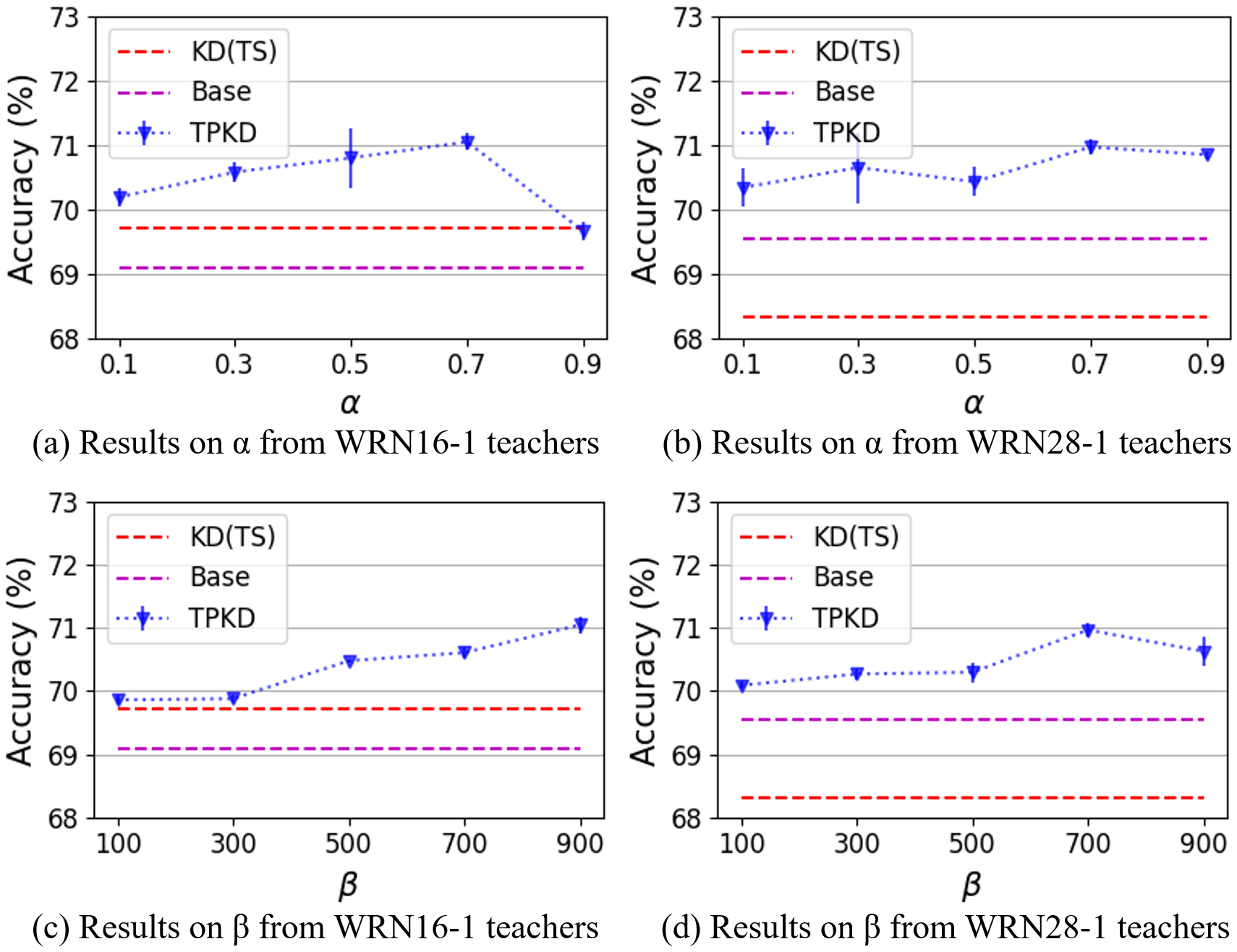}
\caption{Sensitivity to $\alpha$ and $\beta$ of the proposed method for WRN16-1 students on GENEActiv.}
\label{figure:sensitivity}
\end{figure}

%%%%%%%%%%%%%%%%%%%%%%%%
\begin{figure}[htb!]
\centering
\includegraphics[width=3.7in]{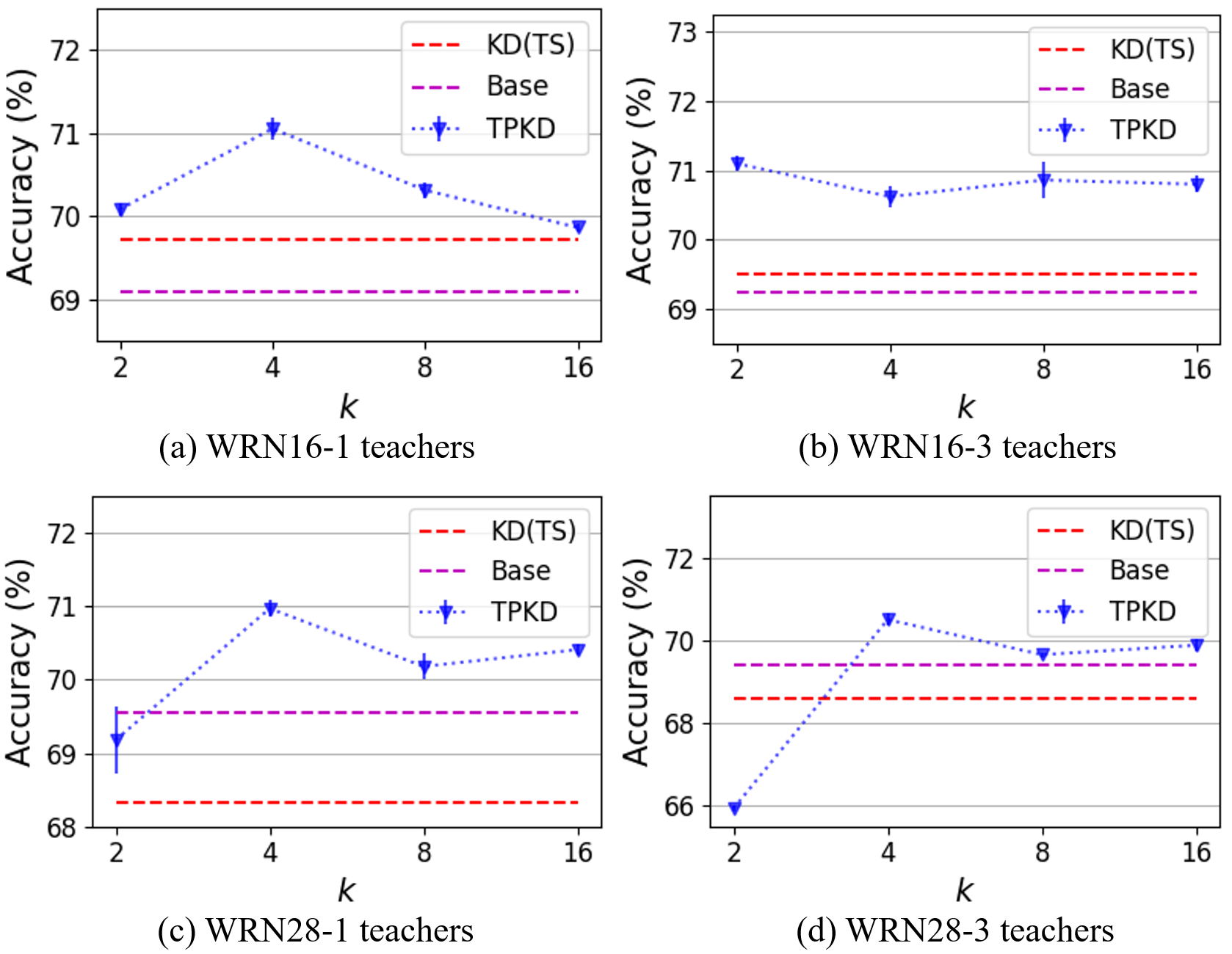}
\caption{Sensitivity to $k$ of the proposed method for WRN16-1 students on GENEActiv.}
\label{figure:k_analysis}
\end{figure}

%%%%%%%%%%%%%%%%%%%%%%%%%%%%%%%%%%%%%%%%%%%%%%%%%%%%
The results of students (WRN16-1), trained with two different teachers by using various $\alpha$ and $\beta$ ($k$ = 4), are illustrated in Figure \ref{figure:sensitivity}. For (a) and (b), $\beta$ is set as the previous section. KD is the result of a student trained with time-series data. Most results from TPKD outperform baselines. The results show their best when $\alpha$ is 0.7. On the other hand, for PAMAP2, their best are shown with $\alpha$ = 0.3. Since GENEActiv has a larger window size and much lower number of channels than PAMAP2, utilizing features from time-series data may help improvements more than PIs. On the other hand, since PAMAP2 has a much smaller window size but more channels, using projected image data from PIs may provide more useful information than raw time-series data. The results with various $\beta$ are shown in (c) and (d) of Figure \ref{figure:sensitivity}. $\alpha$ is set as 0.7. All results from TPKD outperform baselines. The best results are presented with $\beta = 900$ for (c) and $\beta$ = 700 for (b). The majority of the best results in the previous section had beta values of 700 or higher. For PAMAP2, with the same structured teachers, smaller number of $\beta$ (200) shows the best. When the window size is large and the number of channels is small, orthogonal features can have more influence on classification with $\beta \geq 700$. The results of WRN16-1 students with various $k$ are illustrated in Figure \ref{figure:k_analysis}. $\alpha$ is 0.3 and $\beta$ is set as the same for each combination in section \ref{various_capacity}. Most $k$ cases outperform baselines and best result is yielded with $k$ = 4. When teacher models have different width of networks to their student, $k$ = 2 shows lower accuracy than baselines, whereas $k \geq 4$ shows higher one.
And, as described in section \ref{various_capacity} and \ref{various_combt}, most cases on GENEActiv and PAMAP2 perform best when $k$ = 4. Based on these results, setting appropriate hyperparameters has to be considered to generate the best performance.

\subsubsection{Visualization of Models}
%\es{-- visualization of SP maps and pearson correlation of SP map}

\begin{figure*}[htb!] %t!
%\begin{tabular}
\includegraphics[width = 0.92\textwidth]
{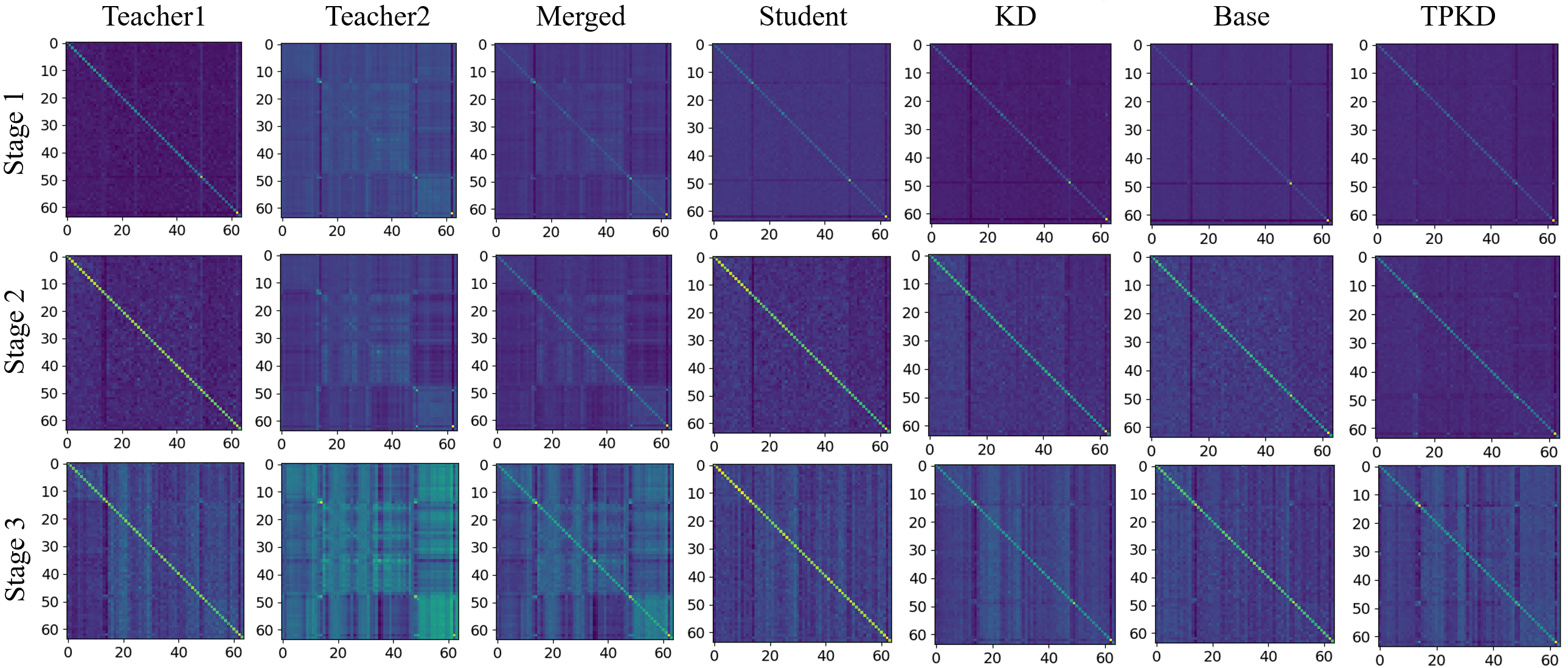} %0.27
\centering
\caption{Activation similarity maps produced by a layer for the indicated stage of the network for a batch on GENEActiv. High similarities for samples of the batch are represented with high values.}
%\end{tabular}
\label{figure:maps_level}
\end{figure*}

\textbf{Analysis of Feature Maps.}
To see more details of activations, we visualize the maps of the teachers (WRN16-3) and student (WRN16-1), representing similarity with high values for inputs. ``Teacher1'' and ``Teacher2'' denote teachers trained with time-series data and PIs, respectively. KD is the result of a student trained with time-series data. Student is the result of a model trained from scratch.
As illustrated in Figure \ref{figure:maps_level}, in all cases, the produced maps in stage 3 have more distinctive patterns than the ones from stage 1 and 2. The maps of two teachers are very different, and the merged one and Student (learned from scratch) are dissimilar, indicating the knowledge gap between them. Some columns of the map from models trained with time-series data are highlighted (Teacher1 and Student). The blockwise patterns are more shown from models trained with PIs. Intuitively, the pattern of the map from Teacher1 is more monotonous than the one from Teacher2. And the diagonal points of the map from models trained with only time-series (Teacher1 and Student) are more prominently highlighted. 
The merged map contains characteristics of both Teacher1 and Teacher2. A student trained with TPKD generates maps closer to those of the merged maps from teachers. The map of a student from TPKD produces brighter colors on diagonal points, which are similar to a merged map, and shows more colorful patterns compared to other students (Student, KD, and Base). Also, the maps from TPKD represent blockwise highlighted features, which verifies that the student preserves topological features by the proposed method.

\begin{figure*}[htb!] %t!
%\begin{tabular}
\includegraphics[width = 0.93\textwidth]
{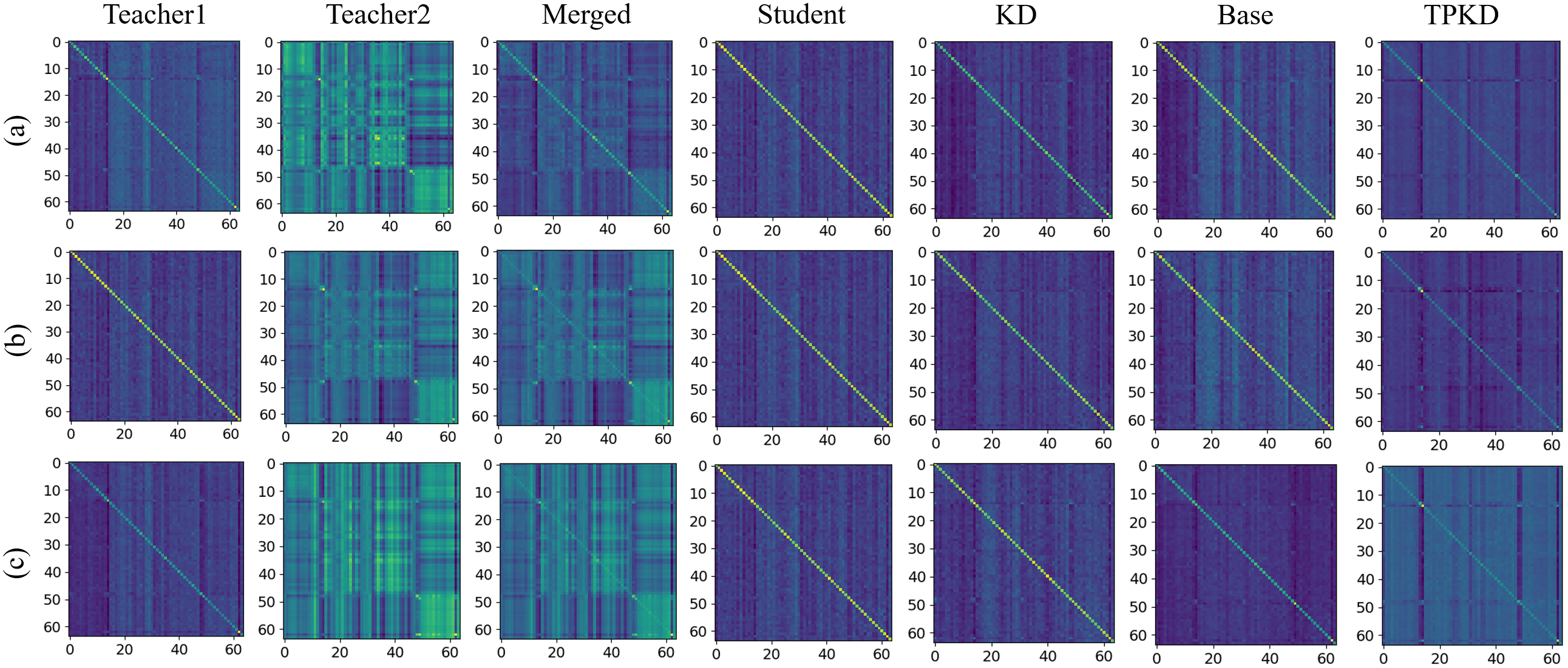} %0.27
\centering
\caption{Activation similarity maps produced by a layer for the stage 3 of the network for a batch on GENEActiv. From (a) to (c), (Teacher1, Teacher2) are (WRN28-1, WRN16-1), (WRN16-1, WRN16-3), and (WRN40-1, WRN28-3), respectively. High similarities for samples of the batch are represented with high values.}
%\end{tabular}
\label{figure:maps_comb}
\end{figure*}

%%%%%%%%%%%%%%%%%%%%%%%%%%%%%%%%%%%%%%%%%%%%%%%%%%%%

More results from different combinations of teachers with a layer for stage 3 of the network are illustrated in Figure \ref{figure:maps_comb}. Compared to baselines, maps of students by TPKD are more similar to the merged ones which contain both topological and time-series features. As illustrated in this figure, compared to baselines, students from TPKD show more blockwise patterns and diagonal points which are similar to a merged maps. In Figure \ref{figure:maps_comb}(c), for TPKD, the contrast between blockwise pattern and monotonous region seems to stand out more than other maps (Figure \ref{figure:maps_comb}(a) and (b)). This combination includes a larger depth and width of Teacher1 and Teacher2 (WRN40-1, WRN28-3) compared to a student (WRN16-1). For (c), the merged map is more different from the map of learning from scratch (Student) than other combinations of teachers. This shows that the knowledge gap increases when the capacity of teachers is much different from that of students. This implies that distilled students from TPKD can acquire features of both Teacher1 and Teacher2 to achieve improved performance even when the knowledge gap increases.
Therefore, TPKD encourages a student to well obtain both features of topological and time-series data while reducing the knowledge gap.

%Therefore, TPKD encourages a student to preserve features of topological as well as time series data while reducing the knowledge gap.

\textbf{Analysis of Model Similarity.} To analyze the similarity of student models generated by different methods, we visualize representations of models with centered kernel alignment (CKA) \cite{raghu2021vision, kornblith2019similarity, cortes2012algorithms} on GENEActiv.
This plot shows similarities between all pairs of layers with different models.

%%%%%%%%%%%%%%%%%%%%%%%%%%%%%%%
\begin{figure*}[htb!]
\centering
\includegraphics[width=6.2in]{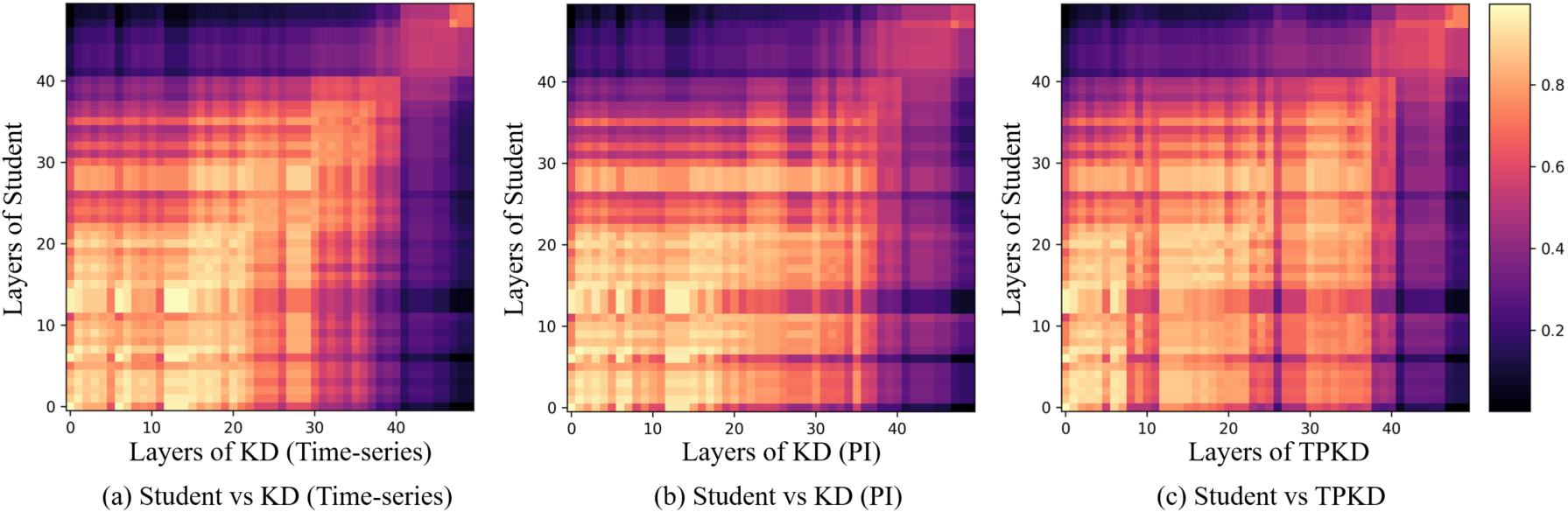}
\caption{Representation of similarities for student models from various methods with all pairs of layers. Teachers are WRN16-3 and students are WRN16-1. ``Student'' denotes a model learned from scratch. Brightness implies similarity scores. High similarity score is represented with high values.}
\label{figure:layersimilarity}
\end{figure*}

As shown in Figure \ref{figure:layersimilarity}, a result of Student (learning from scratch) and KD (time-series) (Figure \ref{figure:layersimilarity}(a)) shows that two models learned similar features in many layers, and the initial layers are more similar than the deeper layers. In Figure \ref{figure:layersimilarity}(b), Student and a student learned with PI have more similarities in lower layers compared to higher layers. For lower layers of similarities, it represents different column-wise patterns and lower values compared to Figure \ref{figure:layersimilarity}(a). For higher layers, it also shows different patterns and less similarity values compared to Figure \ref{figure:layersimilarity}(a). This implies that topological features from PI can provide different features from time-series.

For TPKD (Figure \ref{figure:layersimilarity}(c)), patterns of the representation are much different from results of (a) and (b). After the very first early layers, differences can be observed more prominently with less bright colors in the representation. Also, intuitively, column-wise differences can also be seen.
This represents a distilled student from TPKD and the other baselines do indeed have certain dissimilarities.
Thus, using both topological features and orthogonal properties based on TPKD affects training a student, which provides different features from training with time-series or PI alone.

\subsubsection{Analysis of Orthogonality in Distillation}
To analyze the effects of leveraging orthogonal features, we measure feature similarity quantitatively with Pearson correlation coefficient on activation maps of models from various knowledge distillation methods. Also, we analyze the generalizability of student models for the different methods.

\begin{figure*}[htb!]
\centering
\includegraphics[width=6.2in]{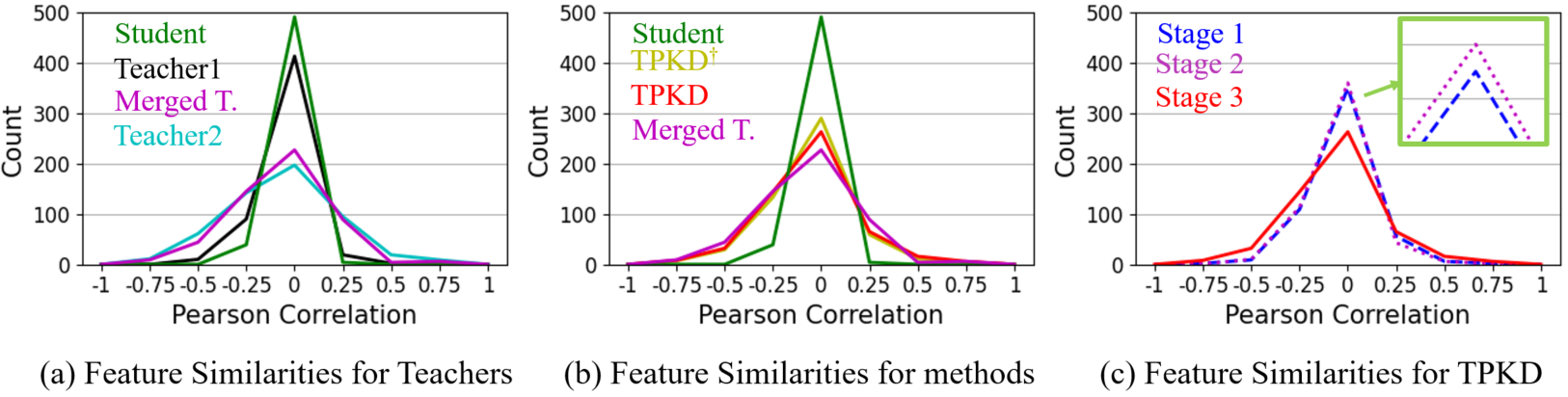}
\caption{Feature similarities for various knowledge distillation methods on GENEActiv. Teachers are WRN28-3 and students are WRN16-1 (1D CNNs). Merged T. denote the merged features from teachers. (a) and (b) are results from 3rd stage of the networks. $\dagger$ denotes without orthogonal features.}
\label{figure:GENE_pcorr}
\end{figure*}

\textbf{Feature Similarity.} We calculate Pearson correlation coefficient on activation similarity maps from intermediate layers. Four patches $\widehat{G} \in \mathbb{R}^{bd \times k}$ = [$\widehat{G}_1$, $\widehat{G}_2$, $\cdots$, $\widehat{G}_k$] ($k$ = 4) from students trained with WRN28-3 teachers are used to generate feature similarity plots. All pair combinations of the patches [($\widehat{G}_1$, $\widehat{G}_2$), ($\widehat{G}_1$, $\widehat{G}_3$), $\cdots$, ($\widehat{G}_{k-1}$, $\widehat{G}_k$)] are considered for the coefficient. As depicted in Figure \ref{figure:GENE_pcorr} (a), the similarities between the two teachers are very different. The model trained from scratch with time-series alone shows high values in 0 of the correlation coefficient. This implies that most of the patches from the models are decorrelated. On the other hand, the patches of Teacher2 are more correlated and much different from Student and Teacher1. The result from the merged patches (Merged T.) for teachers shows intermediate results between two teachers, but closer to Teacher2. These show there is a statistical gap between the teachers and student. In the figure (b), TPKD (with orthogonal features) shows a more similar result to Merged T. than the one without orthogonal features which is a direct map matching method. By orthogonal features in distillation, the student can learn more attentive features and perform more teacher-like tasks. Also, the student trained by TPKD is implemented with time-series data only as an input, but it produces similar features to Merged T. Thus, TPKD distills a student preserving topological features while reducing the knowledge difference between teachers and a student. As described in (c) of the figure, the patches from 3rd stage of the network are more correlated with each other than the other stages. And, the features from each stage have different statistical characteristics. So, transferring features with different stages can help to improve performance.
%%%%%%%%%%%%%%%%%%%%%%%%%%%%%%%

%%%%%%%%%%%%%%%%%%%%%%%%%%%%%%%
%%%%%%%%%%%%%%%%%%%%%%%%%%%%%%%
\textbf{Model Reliability.} To study the generalizability and regularization effects, we measured expected calibration error (ECE) \cite{guo2017calibration} and negative log likelihood (NLL) \cite{guo2017calibration}. ECE is to measure calibration, representing the reliability of the model. The probabilistic quality of a model can be
measured by NLL. We used students trained by teachers of WRN16-3 and WRN28-1. In Table \ref{table:gene}, ECE and NLL with various methods on GENEActiv are described. The results of Base outperform KD and Student (learning from scratch). This implies that leveraging topological features improves performance in reliability. TPKD (with orthogonal features) generates the lowest ECE and NLL in both cases. The results on PAMAP2 are shown in Table \ref{table:pamap}. In both cases, Base performs better than KD and the model learned from scratch. TPKD outperforms all baselines, and using orthogonal features shows the best results. This implies that utilizing orthogonal features in distillation aids in generating a better model, not only for accuracy but also for reliability.

%%%%%%%%%%%%%%%%%%%%%%%%%%%%%%%%%%%%%

\begin{table}[htb!]
\begin{center}
\caption{ECE ($\%$) and NLL for various knowledge distillation methods on GENEActiv. Teachers are WRN16-3 and WRN28-1. Students are WRN16-1 (1D CNNs).}
\label{table:gene}
\scalebox{0.92}{
\begin{tabular}{ c | c  c | c  c }
\hline
\multirow{2}{*}{Method} & \multicolumn{2}{c|}{WRN16-3} & \multicolumn{2}{c}{WRN28-1}\\
 & ECE & NLL & ECE & NLL \\
\hline
Student & 3.548 & 2.067 & 3.548 & 2.067\\

KD & 3.200 & 1.520 & 3.064 & 1.512\\ 

Base & 2.998 & 1.142 & 3.009 & 1.271\\

TPKD (w/o Orth.) & 2.728 & 1.128 & 2.634 & 1.114\\

TPKD (w/ Orth.) & \textbf{2.637} & \textbf{1.103} & \textbf{2.616} & \textbf{1.068}\\
\hline 
\end{tabular}
}
\end{center}
\end{table}

\begin{table}[htb!]
\begin{center}
\caption{ECE ($\%$) and NLL for various knowledge distillation methods on PAMAP2. Teachers are WRN16-3 and WRN28-1. Students are WRN16-1 (1D CNNs).}
\label{table:pamap}
\scalebox{0.92}{
\begin{tabular}{ c | c  c | c  c }
\hline
\multirow{2}{*}{Method} & \multicolumn{2}{c|}{WRN16-3} & \multicolumn{2}{c}{WRN28-1}\\
 & ECE & NLL & ECE & NLL \\
\hline
Student & 2.299 & 1.287 & 2.299 & 1.287 \\

KD & 2.183 & 1.061 & 2.323 & 1.329\\ 

Base & 2.039 & 0.815 & 2.130 & 0.955\\

TPKD (w/o Orth.) & 1.897 & 0.754 & 2.075 & 0.896\\

TPKD (w/ Orth.) & \textbf{1.692} & \textbf{0.708} & \textbf{1.818} & \textbf{0.856} \\
\hline 
\end{tabular}
}
\end{center}
\end{table}

%%%%%%%%%%%%%%%%%%%%%%%%%%%%%%%%%%%%%%%

\subsubsection{Analysis of Invariance to Corruptions}
To explore the model's ability to be robust (and ideally, invariance) to various signal corruptions, we evaluate the models on a noisy testing set, including continuous missing and Gaussian noise, where the corruptions reflect the errors commonly encountered in time-series \cite{jeon2022kd, ijcai2021631, wang2019time}. To consider the case of unknown noise statistics, we set randomly chosen parameters -- ($\kappa_R$, $\sigma_G$) -- denoting the percent of the window size to be removed, and the standard deviation for Gaussian noise respectively. Both corruptions are applied together and we define three levels of corruption; Level 1 (0.15, 0.06), Level 2 (0.22, 0.09), and Level 3 (0.30, 0.12). Note, the classification models were trained with the original training set.

%The exact value for the noise is chosen randomly, which is less than the defined parameter.

\begin{figure}[htb!]
\centering
\includegraphics[width=3.8in]{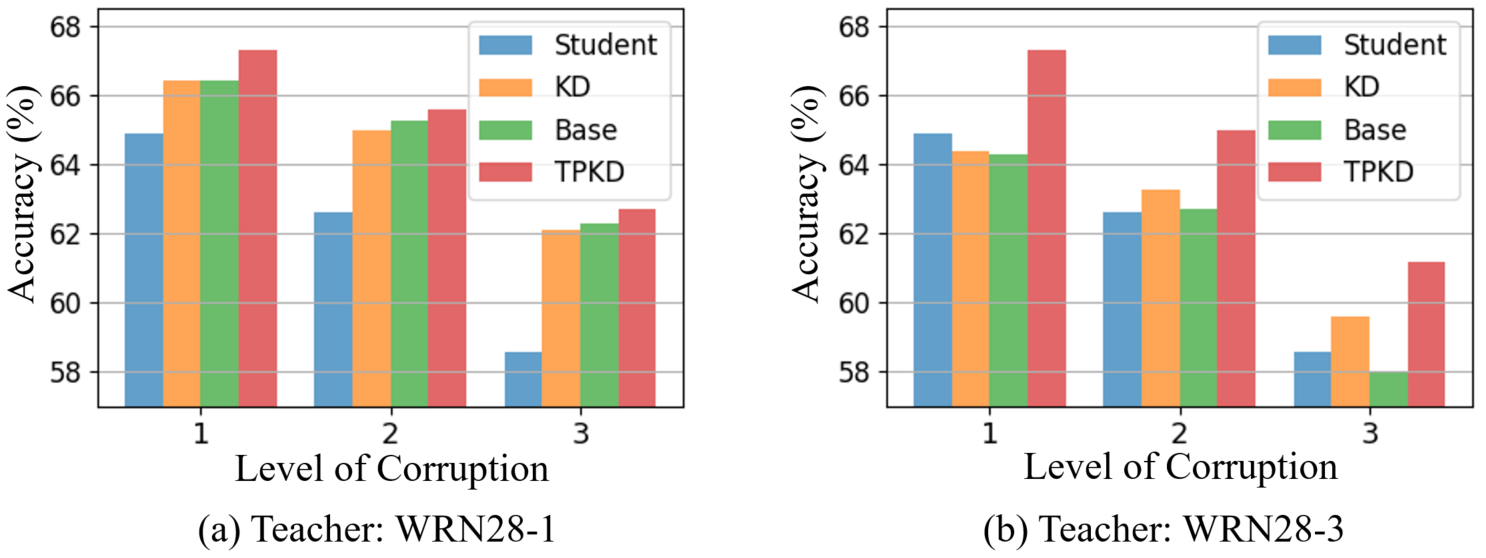}
\caption{Accuracy ($\%$) with various knowledge distillation methods for various signal corruption severity levels on GENEActiv. Students are WRN16-1 (1D CNNs).}
\label{figure:noise_testing}
\end{figure}

As shown in Figure \ref{figure:noise_testing}, TPKD (with orthogonal features) outperforms others in all cases. The results for WRN28-1 teachers show that KD and Base perform better than learning from scratch. The accuracy of Base is higher than KD, which implies that topological features complement the performance. However, when teachers are WRN28-3, there is a case that results of both KD and Base are lower than the model trained from scratch. Even if both models show better performance when testing set is not corrupted, they are sensitive to signal corruptions. Since the capacity of the teacher is much higher than the one of the student, the knowledge difference is larger and it is more difficult to get benefits from distillation. In this case, only TPKD outperforms learning from scratch in all cases. Thus, TPKD helps reducing the knowledge gap to distill a better student.

\subsection{Computational Time}

We compare the computational time of various methods for testing set on GENEActiv. We implemented the evaluation on a desktop with a 3.50 GHz CPU (Intel® Xeon(R) CPU E5-1650 v3), 48 GB memory, and NVIDIA TITAN Xp (3840 NVIDIA® CUDA® cores and 12 GB memory) graphic card \cite{gpuspec}. We tested approximately 6k samples with a batch size of 1. In Table \ref{table:processing_time}, the considered accuracies are the best ones from Table \ref{table:combT_GENE} and \ref{table:mixcombT_GENE}. Since the time is required to generate PIs on the CPU, a model learned from scratch with PIs takes the largest amount of time in the table. A WRN16-1 (1D CNNs) student from TPKD takes the lowest time with the best accuracy. The result on the CPU strongly presents that a model compression method such as KD is required to run on small devices having limited power and computational resources.

%%%%%%%%%%%%%%%%%%%%%%%%%%%%%%

\begin{table}[htb!]
\renewcommand{\tabcolsep}{1.0mm} 
\caption{Processing time of various models on GENEActiv.}
\label{table:processing_time}
\centering
\scalebox{0.75}{
\begin{tabular}{c |c |c |c |c | c}
%\begin{tabular}{|p{7.9em}|p{2.5em} p{3em} p{2.5em} p{3em} p{3em}|} %5.2
\hline
%5em
\multirow{4}{*}{Model} & \multicolumn{2}{c|}{Learning} & \multicolumn{2}{c|}{\multirow{2}{*}{KD}} & TPKD\\

 & \multicolumn{2}{c|}{from scratch} &\multicolumn{2}{c|}{} & (w/ Orth.) \\
  \cline{2-6}
 & TS (1D) & PImage (2D) & TS & PImage & TS+PImage\\
 \cline{4-6}
 & WRN28-3 & WRN16-3 & \multicolumn{3}{c}{WRN16-1 (1D CNNs)} \\
\hline

Accuracy (\%) & 69.23 & 59.8 & 69.71 & 68.76 & \textbf{71.74} \\ \hline %\cline{4-6} 
\multirow{2}{*}{GPU (sec)} & \multirow{2}{*}{29.94} & 356.92 (PIs on CPU) & \multicolumn{3}{c}{\multirow{2}{*}{\textbf{15.23}}} \\
 &  & +13.63 (model) & \multicolumn{3}{c}{}\\ \hline
\multirow{2}{*}{CPU (sec)} & \multirow{2}{*}{1977.89} & 356.92 (PIs on CPU) & \multicolumn{3}{c}{\multirow{2}{*}{\textbf{16.66}}} \\ 
 & & +11191.45 (model) & \multicolumn{3}{c}{}\\
\hline

\end{tabular}
}
\end{table}

%%%%%%%%%%%%%%%%%%%%%%%%%%%%%%%%%%%%

\section{Discussion}
We tested the proposed method, TPKD, with different datasets including different sizes of window length, the number of classes, and complexity. In most of the cases, TPKD outperformed baselines in classification. Even with smaller or the different number of classes, TPKD showed the best accuracy. 
This implies that using orthogonality properties to transfer feature relationship can improve performance significantly. The results also showed that larger capacity of teachers does not guarantee the generation of a better student. This corroborates previous studies \cite{jeon2022kd, choi2018temporal}.
Among different combinations of teachers and students, in most cases, WRN16-3 teachers distilled a superior student. When the window length of a sample is large and the model performs with a smaller number of classes, the smaller network (WRN16-1) showed good performance. This indicates that large-sized models can increase the knowledge gap, and smaller teachers can perform better for easy problems.
For different architectures of teachers, in overall cases, the proposed method generated a better student. However, for WRN28-1 Teacher1 and WRN16-3 Teacher2 on GENEActiv, TPKD without Orth. outperformed with Orth.
This showed that TPKD with Orth. performs better when the width of teachers is similar to a student among various combinations of teachers. Also, even though improvements for large Teacher1 with TPKD are smaller than other combinations, TPKD achieved better results compared to baselines. Therefore, when two teachers are similar or the width of teacher models is similar to a student, the proposed method performs better than the other combinations. And, TPKD can alleviate the negative effects from the knowledge gap in distillation. 

For TPKD, best cases showed with different $\beta$ and $k$ across datasets. In general, when $k$ is 4, TPKD performs the best. However, the optimal $\beta$ is different for datasets. For dataset including 40 channels (PAMAP2), when $\beta$ is 200, it showed best performance in overall cases, whereas it is 700 for GENEActiv with 3 channels. Finding optimal parameters for training can consume time which is a limitation of the method.

The visualized feature maps indicate that distilled students from TPKD can generate topological features, which are shown with blockwise patterns in maps from intermediate layers. Also, the map from the proposed method represents highlighted diagonal points, which are similar to the merged map and Teacher1 learned with time-series. Thus, the student from TPKD includes both characteristics of time-series and topological features from two teachers. Furthermore, similarity comparisons for layers showed that models with time-series and students distilled from TPKD have many differences, which are more represented with lower layers and column-wise patterns.
Therefore, these visualized results show that a student by TPKD can preserve topological features from a teacher model, which cannot be learned with time-series features alone. Also, the results present that utilizing both topological features and orthogonal properties based on TPKD affects KD learning improvement and aids in distilling a superior student than other methods.

\section{Conclusion} \label{sec:conclusion}

In this paper, we propose a framework based on knowledge distillation, for effectively utilizing topological representations of wearable sensor time-series data within deep architectures. This requires reducing the statistical gap between the teacher and student, where we find a beneficial effect of imposing orthogonal constraints between features, further assisted by an annealing strategy. We evaluated the effectiveness of the proposed method, TPKD, under a variety of combinations of KD in classification. TPKD showed more accurate and efficient performance than baselines, which is significant for various applications running on edge devices. 
In future work, we aim to extend the proposed method by leveraging more types of teachers trained with different representations (e.g. Gramian Angular Field based images) of time-series data. Also, we would like to explore the effects of augmentation methods on the representations for using multiple teachers in knowledge distillation.

\section*{Acknowledgements}
This research was funded in part by NIH R01GM135927, as part of the Joint DMS/NIGMS Initiative to Support Research at the Interface of the Biological and Mathematical Sciences, and NIH NIAMS R01AR080826.

%%%%%%%%%%%%%%%%%%%%%%%%%%%%%%%





%\begin{thebibliography}{00}


%\bibitem{}

%\end{thebibliography}
\bibliographystyle{IEEEtran} %elsarticle-num
\bibliography{main}

\end{document}